\documentclass[aps,prl,twocolumn,superscriptaddress,floatfix,longbibliography]{revtex4-1}
\usepackage[utf8]{inputenc}
\usepackage[english]{babel}
\usepackage{amsmath}
\usepackage{amsfonts}
\usepackage{amssymb}
\usepackage{graphicx}
\usepackage{bm}
\usepackage[dvipsnames]{xcolor}
\usepackage[normalem]{ulem}
\usepackage[percent]{overpic}
\definecolor{jadclr}{rgb}{0,0.5,0}
\definecolor{jadcolor}{rgb}{0.5,0,0}
\usepackage[bookmarks=true,colorlinks,linkcolor=jadclr,urlcolor=jadcolor,citecolor=jadcolor]{hyperref}
\usepackage{bbold}

\makeatletter
\DeclareRobustCommand{\cev}[1]{%
  {\mathpalette\do@cev{#1}}%
}
\newcommand{\do@cev}[2]{%
  \vbox{\offinterlineskip
    \sbox\z@{$\m@th#1 x$}%
    \ialign{##\cr
      \hidewidth\reflectbox{$\m@th#1\vec{}\mkern4mu$}\hidewidth\cr
      \noalign{\kern-\ht\z@}
      $\m@th#1#2$\cr
    }%
  }%
}

\begin{document}
\author{Johannes Lang}
\email[]{j.lang@uni-koeln.de}
\affiliation{Institut f\"ur Theoretische Physik, Universit\"at zu K\"oln, Z\"ulpicher Stra{\ss}e 77, 50937 Cologne, Germany}

\author{Subir Sachdev}
\affiliation{Department of Physics, Harvard University, Cambridge MA 02138, USA}

\author{Sebastian Diehl}
\affiliation{Institut f\"ur Theoretische Physik, Universit\"at zu K\"oln, Z\"ulpicher Stra{\ss}e 77, 50937 Cologne, Germany}

\title{Numerical renormalization of glassy dynamics}
\date{\today}

\begin{abstract}
The quench dynamics of glassy systems are challenging. Due to aging, the system never reaches a stationary state but instead evolves on emergent scales that grow with its age. This slow evolution complicates field-theoretic descriptions, as the weak long-term memory and the absence of a stationary state hinder simplifications of the memory, always leading to the worst-case scaling of computational effort with the cubic power of the simulated time. Here, we present an algorithm based on two-dimensional interpolations of Green's functions, which resolves this issue and achieves sublinear scaling of computational cost. We apply it to the quench dynamics of the spherical mixed $p$-spin model to establish the existence of a phase transition between glasses with strong and weak ergodicity breaking at a finite temperature of the initial state. By reaching times three orders of magnitude larger than previously attainable, we determine the critical exponents of this transition. Interestingly, these are continuously varying and, therefore, non-universal. While we introduce and validate the method in the context of a glassy system, it is equally applicable to any model with overdamped excitations.
\end{abstract}
\maketitle

\paragraph{Introduction ---}

Optimizing complex, high-dimensional landscapes is a central challenge in computer science, inspiring influential algorithms~\cite{Kirkpatrick1983, Finnila1994, Earl2005}. This challenge has gained renewed significance with the rise of artificial intelligence, where deep learning models exhibit remarkable generalization despite being massively overparameterized. Yet, why and how neural networks consistently find solutions that generalize well remains one of the most profound open questions in computer science~\cite{Zhang2021}.

In the absence of a fundamental theory, many successful learning strategies remain heuristic, often drawing inspiration from the physics of disordered systems~\cite{Choromanska2015, Chaudhari2019, Baldassi2016, Baity-Jesi2018}. Indeed, it has been shown that stochastic gradient descent, the backbone of deep learning, maps onto the relaxational dynamics of mean-field spin glasses~\cite{Hartnett2018, Mannelli2020, Mignacco2022, Manacorda2022}. This connection highlights the importance of understanding dynamical phase transitions in spin glasses, as jamming transitions are associated with learnability thresholds, while strong ergodicity breaking corresponds to underparameterization, where the network fails to explore all relevant configurations~\cite{Geiger2019, Baity-Jesi2018, Manacorda2022, Mannelli2020}.

While analytical solutions exist for specific mean-field models~\cite{Cugliandolo1993, Cugliandolo1994}, addressing these questions generally requires numerically evolving thermodynamically large systems over extremely long timescales. The efficient simulation of spin glass dynamics is therefore crucial not only for statistical physics but also for advancing our theoretical understanding of deep learning.

The aging dynamics of mean-field spin glasses arises from marginal stability, whereby the evolution in high-dimensional configuration space gets stuck in saddles, not local minima. Consequently, all mean-field spin glasses are (self-organized) critical~\cite{DeDominicis1983,Pazmandi1999,LeDoussal2010}, as seen in the emergent scale invariance of correlations and response functions~\cite{Cugliandolo1995}. The latter is broken on a scale set by the age of the system $t$, making it natural to address aging and the approach to marginal stability from the vantage point of the renormalization group (RG). Although the interaction constant is not renormalized, due to the weak long-term memory characteristic of glassy systems, all coarse features of the correlation and response functions of scale $\sim t$ are relevant, whereas most finer details of the distant past are forgotten and thus irrelevant in the RG sense.

In this work, we introduce an efficient algorithm that exploits this RG flow by systematically discarding irrelevant details of the past through adaptive rescaling of the discretization. Crucially, this rescaling is performed at a fixed cost per time step, ensuring efficiency even for arbitrarily long memory effects. We apply it to the dynamics of the mixed spherical $p$-spin model after a quench to zero temperature from a finite temperature state, resolving the previously inaccessible phase diagram and identifying, for the first time, the critical exponents of the weak-to-strong ergodicity breaking transition.

\paragraph{Keldysh approach to quench dynamics ---}

The time-evolution of any interacting system can be described in terms of correlations $C(t,t')$ and response functions $R(t,t')=\bar{R}(t',t)$. These Green's functions satisfy the Kadanoff-Baym equations~\cite{Kadanoff_book}
\begin{align}\label{eq:EOM}
\begin{split}
    R_0^{-1}\star C=&\Sigma^R\star C+\Sigma^C\star \bar{R}\,,\\
    R_0^{-1}\star R=&\Sigma^R\star R+\delta(t-t')\,.
\end{split}
\end{align}
Here, $R_0$ is the response function of the non-interacting theory and $\Sigma^{C/R}$ are so-called self-energies that can be calculated from two-particle irreducible Feynman diagrams~\cite{Baym1962, Cornwall1974}. Furthermore, $A \star B = \int_{s}A(t,s)B(s,t')$ represents a convolution. For simplicity, we neglect the possibility of spatial dependence. Although formally exact, Eqs.~\eqref{eq:EOM} typically require some approximation for the self-energies. The same problem occurs in equilibrium and can be solved in many cases (see e.g.~\cite{Fetter2003, Altland2010}). Thermalization requires energy redistribution and thus self-energies that are non-local in time. Since  $R_0^{-1}$ can be written as derivative operators, out of equilibrium the equations of motion~\eqref{eq:EOM} take the form of partial integro-differential equations. The computation effort to evolve the system to an age $t$ with a fixed time step and without further approximations scales at least as $t^3$~\cite{Berges2008, Folena2020, Koehler1999, Aoki2014}. 

As glasses break ergodicity, equilibration and accordingly time-translation invariance is dynamically unachievable~\cite{Cugliandolo1995, Bouchaud1992}. Consequently, standard approximations like memory integral truncation~\cite{Stahl2022}, generalized Kadanoff-Baym~\cite{Lipavsky1986}, or the Wigner approximation~\cite{Kamenev_book} are not applicable to glassy dynamics. This calls for an adaptive approach to time discretization--one that captures all relevant scales while remaining computationally efficient.

\paragraph{Dynamical renormalization algorithm ---}
To solve the Kadanoff-Baym equations we aim to evolve two-point functions $A(t,t')$ in time. Even at times much greater than any microscopic scale intrinsic to the system $t\gg \tau_\text{micro}$, they strongly depend on the relative time $\tau=t-t'$ for $\tau\lesssim \tau_\text{micro}$. Furthermore, if the dynamics are initiated by a quench, there may also be a strong dependence on $t'$ for $t'\lesssim \tau_\text{micro}$, regardless of $t$. We therefore parametrize the $A(t,t')$ as $\mathcal{A}(t,\theta)\equiv A(t,\theta t)$ and choose a discretization $t'_i=t \theta_i$, where $\theta_i\in[0,1]$ is a fixed irregular grid of length $N$ that is dense near 0 and 1. The precise choice of the $\theta_i$ is irrelevant as long as 
\begin{align}\label{eq:condition}
    \theta_{i+1}-\theta_i\ll \frac{\tau_\text{micro}}{t} \quad \forall i:\min(\theta_i,1-\theta_i)\lesssim\frac{\tau_\text{micro}}{\min{(t',\tau)}}
\end{align}
for all simulated times $t$. The specific grid used in our calculations, together with some further details of the implementation, are given in the supplemental material. In principle, this grid could be optimized further by making it adaptive. However, this has turned out to be unnecessary for all our simulations, which use an adaptive discretization exclusively of $t$. From the RG perspective, keeping the grid length $N$ fixed under a scale transformation/time evolution $t\to b t$ corresponds to a coarse graining step. Due to condition~\eqref{eq:condition}, it retains sufficient resolution of microscopic scales for $\tau\lesssim \tau_\text{micro}$ and $t'\lesssim \tau_\text{micro}$. Continuing in the RG logic, as the $\theta_i$ are dimensionless, they require no rescaling, which leaves the renormalization step, which is performed by the explicit propagation of the Kadanoff-Baym equations. Aging of classical glasses is then characterized by $C\to C$ and $R\to b^{-1} R$ for $t'\sim t$ and $R\to R$ for $\tau\lesssim \tau_\text{micro}$.

\begin{figure}[!t]
\centering
\includegraphics[width=\columnwidth]{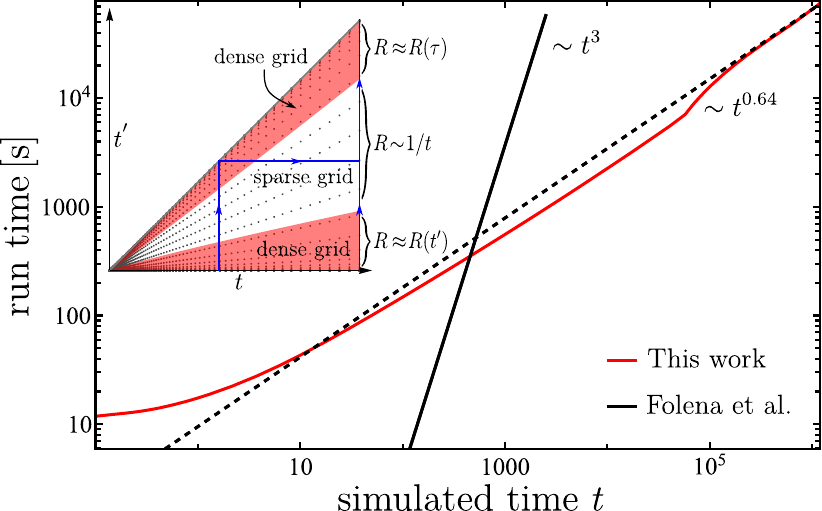}
\caption{Comparison of the performance of the algorithm presented here for the mixed $p$-spin model against a standard implementation of Eqs.~\eqref{eq:EOM_Glass} (see Ref.~\cite{Folena2020}). Both codes are run on comparable modern desktop computers. For the time simulated in the present work, explicit SSPRK methods with adaptive step sizes can be used to reach a sublinear scaling of the computational cost (both memory and CPU time). At even later times implicit methods will be necessary to maintain sublinear scaling. Inset: Visualization of the time discretization and a typical contour of the memory integration in Eq.~\eqref{eq:memory_int} (blue). The shaded red areas highlight regions that are densely sampled (for further details see supplemental material).}
\label{fig:performance}
\end{figure}

In the numerical implementation, time steps in $t$ are chosen adaptively by imposing an upper bound on the error estimate of the ODE solver used to propagate Eqs.~\eqref{eq:EOM} forward in time. In the specific cases considered below, stability requires strong stability preservation (SSP). Hence, at late times, we use an explicit fourth-order SSP Runge-Kutta method with optimal SSP time-step restrictions known as SSPRK(10,4)~\cite{Ketcheson2008, Fekete2022}. A depiction of the resulting grid is shown in the inset of Fig.~\ref{fig:performance}.

The main challenge resulting from the (near) optimal time discretization is the increased difficulty of computing memory integrals. As can be seen in the inset of Fig.~\ref{fig:performance}, integration over the past requires interpolation in an unstructured two-dimensional grid. Moreover, the interpolation order should be at most one order lower than that of the ODE solver. In practice, as the necessary time derivatives are anyway calculated for the forward propagation in time $t$, it is straightforward to implement a third-order Hermite interpolation.

Using the symmetries of the Green's functions, memory integrals take the form
\begin{align}\label{eq:memory_int}
\begin{split}
    \int_0^t \!ds A(t,s)B(s,t')=&t\underbrace{\int_0^\theta \!d\phi \mathcal{A}(t,\phi)\mathcal{B}(\theta t,\phi/\theta)}_{=I_1(t,\theta)}\\&+t\underbrace{\int_\theta^1 \!d\phi \mathcal{A}(t,\phi)\mathcal{B}(\phi t, \theta/\phi)}_{=I_2(t,\theta)}\,.
\end{split}
\end{align}
When evaluating these integrals numerically, it is crucial that at each step of the memory integration the discretization of $\phi$ satisfies condition~\eqref{eq:condition} while using a fixed number of sampling points. We achieve this by defining $\phi^{(1)}_{ij}=\theta_i\theta_j$, $\phi^{(2)}_{ij}=\theta_j+(1-\theta_j)\theta_i$ and discretizing the integrals as follows
\begin{align}
\begin{split}
    I_1(t_n,\theta_j)&=\sum_i w_i \mathcal{A}(t_n,\phi^{(1)}_{ij})\mathcal{B}(\theta_j t_n,\theta_i)\\
    &=\sum_{ik} w_i M^{(1)}_{ijk}\mathcal{A}(t_n,\theta_k)\mathcal{B}(\theta_j t_n,\theta_i),\\
    I_2(t_n,\theta_j)&=\sum_i w_i \mathcal{A}(t_n,\phi^{(2)}_{ij})\mathcal{B}(\phi^{(2)}_{ij}t_n,\frac{\theta_j}{\phi^{(2)}_{ij}})\\
    &=\sum_{ikl}w_i M^{(2)}_{ijk}\mathcal{A}(t_n,\theta_k)M^{(3)}_{ijl}\mathcal{B}(\phi^{(2)}_{ij}t_n,\theta_l)\,.
\end{split}
\end{align}
Here, $w_i$ are the weights of a quintic spline interpolation. These, as well as the interpolation coefficients $M^{(1,2,3)}_{ijk}$ (we use a high-order Hermite interpolation), can be calculated during the initialization because $\theta$ and $\phi^{(1,2)}$ are fixed.
Since the $M^{(1,2,3)}_{ijk}$ are sparse, computing $\mathcal{A}(t_n,\phi_{ij})$ etc. takes constant time.

What is left is the interpolation of the first argument of $\mathcal{B}$. Since the $t$-grid is adaptive, this needs to be repeated in each time step. In practice, however, the time step $\delta t$ is much smaller than $t$. Hence, initializing an interpolation search with the previous step’s result determines the interpolation coefficients $T^{(1,2)}$ in constant time. For the application to the spherical mixed $p$-spin model, we use a cubic Hermite interpolation.

The final expressions for the memory integrals read
\begin{align}
\begin{split}
    I_1(t_n,\theta_j)&=\sum_{ikm} w_i M^{(1)}_{ijk}\mathcal{A}(t_n,\theta_k) T^{(1)}_{jm}\mathcal{B}(t_m,\theta_i),\\
    I_2(t_n,\theta_j)&=\sum_{iklm} w_i M^{(2)}_{ijk}\mathcal{A}(t_n,\theta_k) M^{(3)}_{ijl} T^{(2)}_{ijm} \mathcal{B}(t_m,\theta_l)\,.
\end{split}
\end{align}
Due to the sparseness of all interpolation tensors, the computational overhead of the algorithm is minimal. The time-critical step is the semi-random memory access to stored correlation and response functions, not determining the interpolation coefficients. Each time step requires $\mathcal{O}(N^2)=\text{const.}$ memory accesses. Hence, each step has a fixed computational cost. Combined with the adaptive $t$-grid, this achieves sublinear scaling for all simulated times (see Fig.~\ref{fig:performance}).

A different approach, related in spirit, has previously been developed by Kim and Latz~\cite{Kim2001, Berthier2007}. In their approach, the time grid contains a fixed number of points with constant spacing. Once the grid is filled, every second point is removed, and the evolution continues with twice the time step. This procedure is highly efficient but relies on a rigid, iterative contraction of the grid, whose stability is not controlled. Indeed, it fails for the spherical mixed 
$p$-spin model, where errors diverge rapidly~\cite{Folena2020, Folena2020b}. The problem arises because an equidistant grid cannot adequately capture all RG-relevant memory effects. The dynamical renormalization algorithm avoids this by selecting suitable $\theta_i$ and using adaptive time steps.

\paragraph{Application: Mixed \texorpdfstring{$p$}{TEXT}-spin model ---}

We now apply the method described above to the mixed $p$-spin model~\cite{Barrat1997, Crisanti2004, Crisanti2006}
\begin{align}
\begin{split}
    H&=\sqrt{\lambda} H_p+\sqrt{1-\lambda} H_s+\sum_i h_i\sigma_i\\
    H_p&=\sum_{i_1,i_2,\dots,i_p}J_{i_1i_2\dots i_p}\sigma_{i_1}\sigma_{i_2}\dots\sigma_{i_p}\,,
\end{split}
\end{align}
where $\sigma_i$ are $M$ real variables subject to the spherical constraint $\sum_{i=1}^M \sigma_i^2=M$, $J_{i_1i_2\dots i_p}$ are Gaussian variables with zero mean and variance $1/(2M^{p-1}p!)$ and the external field is set to zero $h_i=0$. This model occurs naturally in the mode-coupling theory of supercooled liquids~\cite{Goetze1992} as well as optimization problems~\cite{Monasson1997}. At times before $t=0$, the system is in equilibrium with a bath at temperature $T$. For simplicity, we set the system-bath coupling to unity, which together with the variance of the interaction strength fixes $\tau_\text{micro}=\mathcal{O}(1)$. At $t=0$, the bath temperature is quenched to zero. We then study the subsequent relaxation dynamics as a function of the initial temperature $T$, while keeping the final temperature fixed at zero. The pronounced sensitivity of the late-time behavior to the starting temperature is a hallmark of glassy dynamics.
To characterize the relaxation dynamics, we analyze the evolution of two key observables: the correlation function $C(t,t')=\langle\sum_{i=1}^M\sigma_i(t)\sigma_i(t')\rangle/M$
and the response function $R(t,t')=\langle\sum_{i=1}^M d \sigma_i(t)/d h_i(t')\rangle/M$, which satisfy the coupled integro-differential equations~\cite{Franz1995b}
\begin{widetext}
\begin{align}\label{eq:EOM_Glass}
\begin{split}
    \partial_t C(t,t')&=-\mu(t)C(t,t')+\int_0^t ds f''(C(t,s))R(t,s)C(s,t')+\int_{0}^{t'}ds f'(C(t,s))R(t',s)+f'(C(t,0))C(t',0)/T,\\
    \partial_t R(t,t')&=-\mu(t)R(t,t')+\int_{t'}^t ds f''(C(t,s))R(t,s)R(s,t'),\\
    \mu(t)&=\int_0^t ds \left[f''(C(t,s))C(s,t)+f'(C(t,s))\right]R(t,s)+f'(C(t,0))C(t,0)/T\,.
\end{split}
\end{align}
\end{widetext}
Here, $\mu(t)$ ensures the spherical constraint $C(t,t)=1$ and $f(x)=\lambda x^p + (1-\lambda) x^s$ with $f'(x)\equiv \partial_x f(x)$. From the correlation and response functions, all observables can be calculated, for example, the energy of the system is given by
\begin{align}
    E(t)=-\int_0^t ds f'(C(t,s))R(t,s)-f(C(t,0))/T\,.
\end{align}

While the late-time dynamics of the pure $p$-spin model $(p=s)$ are fully understood~\cite{Cugliandolo1993, Arous2006}, the mixed model presents a richer phenomenology already in equilibrium~\cite{Auffinger2022} and its dynamics remains an active area of research~\cite{Cugliandolo2023}. Recently, Folena et al.~\cite{Folena2020, Folena2023} have found indications that for integers $s>p>1$, the system remembers its initial configuration, casting doubt on the paradigm of spin glass dynamics: weak ergodicity breaking. The latter assumes that a glassy system is unable to explore the entire configuration space in finite time~\cite{Bouchaud1992}, but simultaneously deviates infinitely far from the initial state, which implies $q_0\equiv\lim_{t\to\infty}C(t,0)=0$. However, due to limitations in the accessible times, it was not possible so far to resolve whether strong ergodicity breaking, for which the system keeps overlap with the initial state, {\it i.e.\/} $q_0\neq 0$, is ubiquitous in the mixed $p$-spin model or occurs only below some critical temperature. Using the method above, we are now able to resolve this question, establishing the existence of a finite critical temperature.

To numerically test the presence of strong ergodicity breaking, we perform simulations on a standard desktop computer with a fixed grid length of $N=256$. We also conducted finite-size scaling up to 
$N=1024$ to ensure that the presented results are free of finite-size effects. 
We emphasize that due to error propagation through the memory integrals, the evolution to later times requires higher accuracy per time step. High-order Runge-Kutta methods mitigate this, enabling longer simulations without affecting scaling. We impose an error bound of $2*10^{-11}$ on the 1-norm $||C(t,\theta_i t)||_1+||R(t,\theta_i t)||_1$ per step.

At temperatures below the lower critical temperature $T_{c_l}=f'(q_{c_l})/(q_{c_l}\sqrt{f''(1)})$, where $q_0=q_{c_l}=\sqrt{1-f'(1)/f''(1)}$, the system relaxes into a local minimum in configuration space that has a finite overlap with the initial state. Consequently, equilibration occurs without aging to an asymptotic stationary solution of the equations of motion~\eqref{eq:EOM_Glass} that can be found analytically. Since this phase exhibits no glassiness but long-range order in time, we refer to it as ferromagnetic (FM). For pure $p$-spin models $T_{c_l}$ coincides with the mode-coupling temperature $T_\text{MCT}$~\cite{Goetze2008,Folena2020}.

For $T_{c_l}<T<T_{c_u}$, we find what we refer to as a \emph{strong glass} in which the system strongly breaks ergodicity by maintaining some memory of the initial state. In configuration space, the evolution is therefore constrained to the neighborhood of the initial state~\cite{Folena2020}. As opposed to the low-temperature phase, however, this neighborhood is large enough to include an infinite number of saddle points or marginal states that slow down the evolution indefinitely, implying aging. Correspondingly, the system satisfies the marginality condition $\mu(t\to\infty)=\mu_{M}=2\sqrt{f''(1)}$~\cite{Cavagna1998}, while the energy drops below that of a weak glass $E(t\to\infty)< E_W=\sqrt{f''(1)} \left(\frac{f(1)-f'(1)}{f''(1)}-\frac{f(1)}{f'(1)}\right)$. 

\begin{table}
\centering
\begin{tabular}{ |c|c|c|c|c|c|  }
 \hline
 $p$ & $s$ &$\lambda$&$T_{u_2}$&$\nu$ &$\eta$\\
 \hline
 \hline
 3   & 4    &1/2   &$1.0377\pm0.0002\, T_\text{MCT}$& $1.95\pm0.05$ & $0.49\pm0.02$\\
 3   & 4    &3/4   &$1.06106\pm0.00003\, T_\text{MCT}$& $1.85\pm0.03$ & $0.454\pm0.005$\\
 3   & 5    &1/2   &$1.100\pm0.001\, T_\text{MCT}$& $1.88\pm0.05$ & $0.36\pm0.02$\\
 3   & 6    &1/2   &$1.1493\pm0.0003\, T_\text{MCT}$& $1.93\pm0.03$ & $0.291\pm0.006$\\
 3   & 9    &1/2   &$1.254\pm0.001\, T_\text{MCT}$& $1.44\pm0.06$ & $0.36\pm0.04$\\
 2   & 5    &1/2   &$0.9230\pm0.0002$& $1.58\pm0.04$ & $0.67\pm0.02$\\
 \hline
\end{tabular}
\caption{Table of critical temperatures and critical exponents for various mixed $p$-spin models.}\label{tab:critData}
\end{table}

\begin{figure}[!t]
\centering
\includegraphics[width=\columnwidth]{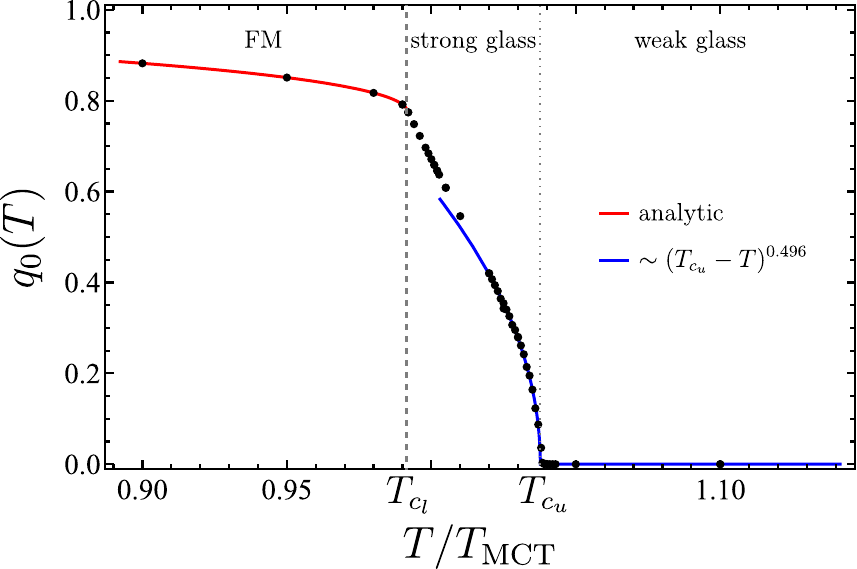}
\caption{Asymptotic initial state correlations $q_0$ as a function of the initial temperature $T$ in units of the mode coupling temperature for the mixed $p$-spin model with $p=3$, $s=4$, and $\lambda=1/2$. Below the lower critical temperature $T_{c_l}$, no aging is observed, and the system remains strongly correlated with the initial state. Above the upper critical temperature $T_{c_u}$, one finds a weak long-term memory and weak ergodicity breaking. The strong glass phase between the critical temperatures is characterized by the coexistence of aging and strong ergodicity breaking. The blue line is a numerical fit to the critical behavior near $T_{c_u}$ from which we extract the critical exponent $\eta$. The red line represents the analytic result for the low-temperature phase.}
\label{fig:critExp}
\end{figure}

Explaining the strong glass in configuration space requires information about the connectedness of the latter, making purely combinatorial arguments insufficient to predict the critical temperature~\cite{Folena2020}. Due to the explicit dependence of the asymptotic evolution on early quench dynamics, we lack an analytical solution for the glass phase. Instead, we extract the critical exponents from the evolution to $t=10^6$. Defining the reduced temperature $\vartheta=(T-T_{c_u})/T_{c_u}$, we find the critical behavior of the initial state correlations $q_0(\vartheta<0)\sim |\vartheta|^\eta$, and energy, $E_W-E(\vartheta<0)\sim |\vartheta|^\nu$. The temperature dependence of $q_0$ for the representative model with $p=3$, $s=4$, and $\lambda=1/2$ is shown in Fig.~\ref{fig:critExp}. Table~\ref{tab:critData} lists the critical temperature and exponents for various mixed $p$-spin models. A strong glass phase was found for all mixed $p$-spin models we investigated. However, the exponents vary continuously, realizing the rare scenario of nonuniversal criticality~\cite{Baxter1971, Kadanoff1971, Bernardi1995, Jin2012}. Notably, the effective temperature of the strong glass is a non-monotonic function of correlations, complicating solutions via the Cugliandolo-Kurchan ansatz~\cite{Cugliandolo1993, Cugliandolo1994, Cugliandolo1995} (see supplemental material).

Finally, at high initial temperatures $T>T_{c_u}$, the system forms what we call a \emph{weak glass}, which is characterized by conventional weak ergodicity breaking and weak long-term memory consistent with the solution of Cugliandolo and Kurchan~\cite{Cugliandolo1995}. It satisfies the marginality condition, just like the strong glass, but also the energy relaxes to the threshold value $E(t\to\infty)=E_W$. For the parameter sets in Table~\ref{tab:critData}, we find a constant effective temperature $x=\sqrt{f''(1)}/f'(1)-1/\sqrt{f''(1)}$, consistent with one-step replica symmetry breaking (1-RSB) in equilibrium~\cite{Lang2024b}.

Although not shown here, the transition between weak and strong glass has been found in all mixed spherical $p$-spin models we investigated. It is, however, absent in all pure spherical $p$-spin models, where the transition occurs directly from the weak glass to a ferromagnetic phase at the mode-coupling temperature $T_c=T_\text{MCT}$.

\paragraph{Outlook --- }\label{sec:Outlook}
We have introduced a new efficient algorithm for the evolution of the Kadanoff-Baym equations for systems with slow dissipative dynamics. We have employed this approach to calculate the transition temperature and the critical exponents of the previously inaccessible weak-to-strong glass transition of the mixed $p$-spin model. 

When applied to spin glasses, our method asymptotically, {\it i.e.\/} at times later than simulated here, approaches linear scaling in computational cost with simulation time. This behavior arises because the equations of motion are stiff, ill-conditioned partial integro-differential equations. Although the system evolves slowly, absolute stability constrains the step size of explicit Runge-Kutta methods. Since the Jacobian can be computed efficiently, strong stability-preserving implicit methods~\cite{Ketcheson2009, Izzo2022} would further improve the asymptotic scaling. Additionally, the current implementation is limited by random memory access latencies, suggesting that substantial performance gains could be achieved through an implementation on graphics processing units~\cite{Bernaschi2020}.

Finally, we emphasize that the algorithm developed in this work is highly generalizable. It naturally extends to other spin glass models, including the classical and quantum Sherrington-Kirkpatrick models~\cite{Kennett2001b, Bernaschi2020}. Furthermore, since the self-energies in the Kadanoff-Baym equation~\eqref{eq:EOM} can be obtained from impurity solvers, the numerical renormalization algorithm presented here can be directly applied to analyze slow modes in dynamical mean-field theory~\cite{Aoki2014}. A promising application is studying neural network learning dynamics via stochastic gradient descent, which map to Kadanoff-Baym equations of disordered mean-field models~\cite{Mannelli2020, Mignacco2022}. Understanding the Gardner transition~\cite{Gardner1987} and strong ergodicity breaking in this context could yield insights into training performance and generalization in artificial neural networks~\cite{Choromanska2015, Baity-Jesi2018}. With the growing interest in statistical physics-inspired approaches to deep learning~\cite{Dean2007, Choromanska2015,Baity-Jesi2019}, our method provides a powerful tool for exploring these connections and advancing both fields.

\paragraph{Acknowledgements --- }
The work of J.L. and S.D. was supported by the Deutsche Forschungsgemeinschaft (DFG, German Research Foundation) under Germany’s Excellence Strategy Cluster of Excellence Matter and Light for Quantum Computing (ML4Q) EXC 2004/1 390534769, and by the DFG Collaborative Research Center (CRC) 183 Project No. 277101999. S.S. was supported by the U.S. National Science Foundation grant No. DMR-2245246.

After this work was completed, a related study employing a different integration scheme that achieves comparable time scales was published~\cite{Citro2025}.

\newpage
\bigskip
\pagebreak
\widetext
\begin{center}
\textbf{\large --- Supplemental Material ---\\ Numerical renormalization of glassy dynamics}\\
\medskip
\text{Johannes Lang, Subir Sachdev, and Sebastian Diehl}
\end{center}

\setcounter{equation}{0}
\setcounter{figure}{0}
\setcounter{table}{0}
\makeatletter
\renewcommand{\theequation}{S\arabic{equation}}
\renewcommand{\thefigure}{S\arabic{figure}}
\renewcommand{\bibnumfmt}[1]{[S#1]}
\renewcommand{\citenumfont}[1]{S#1}
\newcommand{\rmf}{{\rm f}}

\section{Details of the numerical implementation}

The algorithm presented in the main text is largely independent of specific parameter choices. Nevertheless, to ensure full reproducibility, we provide here the details of the specific choices used in the numerical simulations, including our strategy for reducing the memory footprint.

The discretization for the time ratio $\theta = t'/t$ must be dense near $\theta = 0$ and $\theta = 1$. In our simulations, we used the following grid
\begin{align}\label{eq:theta}
    \theta_i=\frac{\tan ^{-1}\left(e^{\alpha }\right)-\tan ^{-1}\left(e^{-\alpha y_i}\right)}{\tan ^{-1}\left(e^{\alpha }\right)-\tan
   ^{-1}\left(e^{-\alpha }\right)}\,,
\end{align}
where $\alpha=-W_{-1}\left(-1/t_\text{max}\right)$ is the Lambert $W$ function with $t_\text{max}$ the largest simulated time (typically $10^6$) and $y_i=(2i-1-N)/(N-1)$ with $i=1,\dots,N$. Eq.~\eqref{eq:theta} has the convenient (although unnecessary) property of being analytically invertible. Depending on the model parameters, more aggressive grids with fewer data points may achieve the same accuracy. However, in our experience, these performance gains are relatively small ($\lesssim 50\%$). The grid presented here is deliberately chosen to be on the conservative side, ensuring controlled errors for all simulated model parameters.

\begin{figure}[htp]
\centering
\includegraphics[width=0.9\columnwidth]{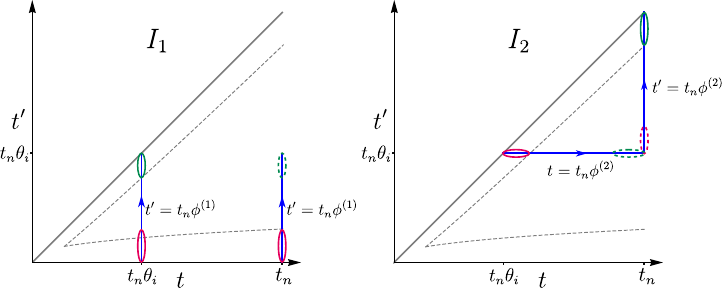}
\caption{Contours on which the integrands of $I_1$ and $I_2$ are evaluated. The choice of discretization in terms of $\phi^{(1,2)}$ is indicated. Red and green ellipses denote areas requiring dense sampling. The need for commensurate sampling of $t$ and $t'$ leads to dense sampling of the regions highlighted by dashed ellipses.}
\label{fig:contours}
\end{figure}

The memory integration requires evaluation of the integrals $I_1$ and $I_2$ defined in the main text and visualized in Fig.~\ref{fig:contours}. It is straightforward to verify that the contours defined by $\phi^{(1)}_{ij} = \theta_i \theta_j$ and $\phi^{(2)}_{ij} = \theta_j + (1 - \theta_j) \theta_i$ are dense in precisely the regions needed for accurate sampling of the integrands.

For most stiff ordinary differential equations, the performance of adaptive solvers strongly depends on the choice of step sizes—a topic that has been extensively studied. Today, well-tested strategies based on control theory principles \cite{Gustafsson1994S,Gustafsson1997S,Burrage2004S} are widely implemented. The aging dynamics of spin glasses, however, are relatively simple in this respect. Because the evolution lacks sudden shocks, even basic step-size control strategies perform well. In our implementation, we adopt a simple strategy: if the error estimate from the embedded Runge-Kutta method satisfies $||C(t, \theta_i t)||_1 + ||R(t, \theta_i t)||_1 < \delta$ with $\delta = 10^{-11}$, the step size is increased by a factor of $1.01$. Conversely, if the estimated error exceeds $2\delta$, the next step size is scaled by $0.9$.  In practice, the step size never decreases at late simulated times.

For most of the computations, we use the fourth-order strong stability strong-stability-preserving Runge-Kutta method SSPRK(10,4). This choice is made to optimize the maximal time step. At shorter times, however, it is more efficient to use higher order methods that lack strong stability preservation. In our implementation, we use the adaptive Dormand-Prince method \cite{Dormand1980S}, which is a six-step, fifth-order method. Due to its higher order it allows for larger time steps at short times. However, its region of stability along the negative real axis only extends to $-3.307$. We estimate the upper bound of the spectral radius of the Jacobian $\rho< 4\sqrt{f''(1)}$. Once the time step $\Delta t$ satisfies $\rho \Delta t>3$, we switch to the slower, yet more stable, SSPRK(10,4). Following the switch, we set $\Delta t\to \Delta t/2$ to account for the lower order of the method, after which it is updated according to the rule described above. The increased number of stages (and therefore integral evaluations) in SSPRK(10,4), combined with the temporary reduction in step size, accounts for the cusp in the performance plot shown in the main text. This occurs because the switch is made slightly too early, causing a minor drop in efficiency. However, the overall impact is negligible.

We have experimented with other explicit Runge-Kutta methods featuring extended stability domains, including SERK2 \cite{Vaquero2009S}, ESERK \cite{Vaquero2016S}, and ROCK4 \cite{Abdulle2002S}. However, their lack of strong stability preservation ultimately limits the maximum step size, offering no performance benefit over SSPRK(10,4).

Despite the use of high-order adaptive methods, the small per-step error bound leads to a large number of time steps overall. For simulations up to $t_\text{max} = 10^6$, we require approximately $10^6$ steps. For the finest discretization of $\theta$ with $N = 1024$, the full history of correlation and response functions therefore occupies roughly $32~\text{GB}$ of memory. While this memory load has little effect on runtime, it approaches the limits of standard desktop systems. History data of the Green's functions only needs to be kept if it is not adequately approximated by the cubic Hermite interpolations $T^{(1,2)}$. Consequently, we use a cubic Hermite interpolation of its left and right neighbors to check the necessity of retaining a time step in the history. If the error between interpolation and history data $||C(t,\theta_i t)||_1+||R(t,\theta_i t)||_1< 10^{-11}$ the time step is erased from the memory. This history pruning is performed every $10^5$ time steps and typically reduces peak memory usage by a factor of about three.

\section{Effective temperature of the strong glass phase}

Thermal equilibrium is characterized by the fluctuation-dissipation relation (FDR), which links correlations and response functions. For classical systems, it takes the form  
\begin{align}
    R(t,t')=\beta \frac{d C(t,t')}{dt'}\,,
\end{align}  
where $\beta$ is the inverse temperature of the equilibrium state. All ergodic systems in contact with a thermal bath eventually equilibrate to the bath's temperature. However, glasses are non-ergodic and never reach thermal equilibrium. Instead, their (unreachable) stationary state exhibits replica symmetry breaking. Correspondingly, glasses satisfy a generalized fluctuation-dissipation relation (gFDR)
\begin{align}
    R(t,t')=x(t,t') \frac{d C(t,t')}{dt'}\,,
\end{align}  
where the inverse effective temperature $x(t,t')$ does not asymptotically converge to $\beta$ but can instead be expressed as a function of the correlations, $x(t,t')=x(C(t,t'))$~\cite{Cugliandolo1994S,Cugliandolo1995S}.  

\begin{figure*}[htp]
\centering
\includegraphics[width=0.45\columnwidth]{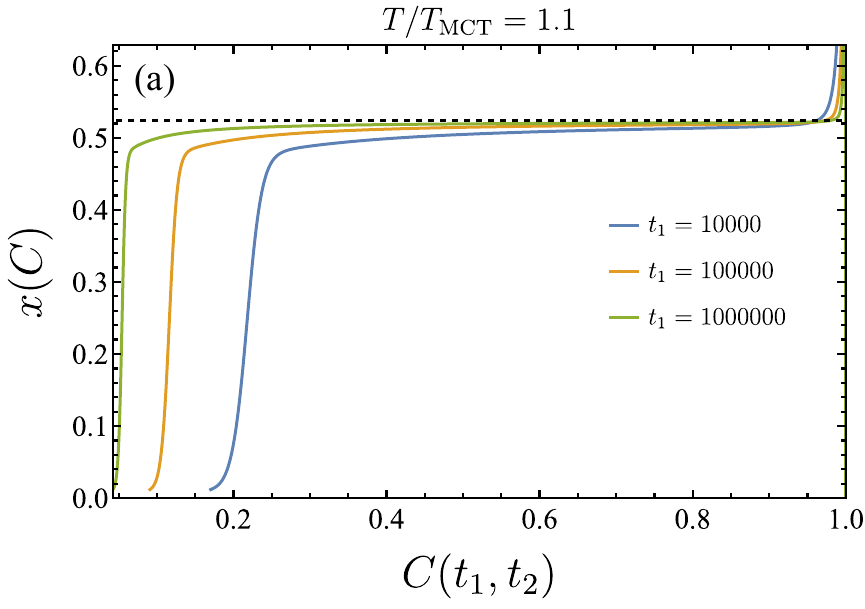}\quad
\includegraphics[width=0.45\columnwidth]{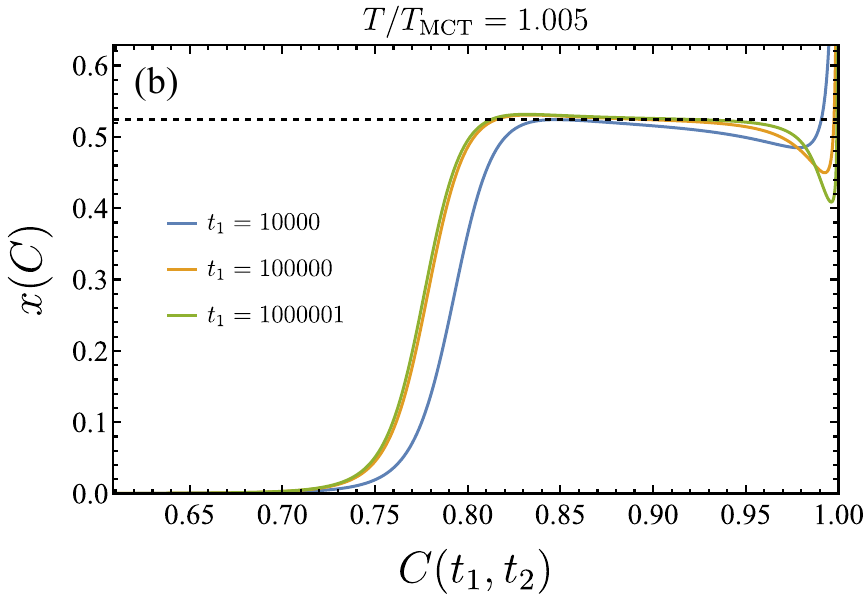}
\caption{(a) Inverse effective temperature at various times for $p=3$, $s=4$, and $\lambda=1/2$ in the weak glass phase. As the system's age increases, the effective temperature approaches a constant. The dashed line indicates the marginal value given by Eq.~\eqref{eq:x}. (b) In the strong glass phase, the effective temperature remains non-monotonic at all simulated times.}
\label{fig:x_comp}
\end{figure*}

Up to an invariance under time reparametrizations, $t\to h(t)$, the asymptotic dynamics are fully characterized once $x(C)$ is determined~\cite{Cugliandolo1993S, Cugliandolo1994S, Cugliandolo1995S}. In the case of the pure $p$-spin model, this was famously achieved by Cugliandolo and Kurchan~\cite{Cugliandolo1993S} under the assumption that $x(C)$ is an increasing function of the correlations. However, for the mixed $p$-spin model, $x(C)$ is generally unknown.  

Here, we find that in the weak glass phase, the effective temperature converges to the marginal constant  
\begin{align}\label{eq:x}
x=\sqrt{f''(1)}/f'(1)-1/\sqrt{f''(1)}  
\end{align}  
consistent with previous results~\cite{Folena2020S}. This behavior is shown in Fig.~\ref{fig:x_comp}(a) and should be compared with the strong glass case in Fig.~\ref{fig:x_comp}(b). There, we find that $x(C)$ converges to a non-monotonic function, which negates a central assumption of the methodology pioneered by Cugliandolo and Kurchan~\cite{Cugliandolo1995S}. This finding offers a possible explanation for the difficulties in developing a consistent ansatz for the effective temperature of the strong glass~\cite{Folena2020S, Folena2023S} and provides guidance for future Ansätze.


\begin{thebibliography}{61}
\expandafter\ifx\csname natexlab\endcsname\relax\def\natexlab#1{#1}\fi
\expandafter\ifx\csname bibnamefont\endcsname\relax
  \def\bibnamefont#1{#1}\fi
\expandafter\ifx\csname bibfnamefont\endcsname\relax
  \def\bibfnamefont#1{#1}\fi
\expandafter\ifx\csname citenamefont\endcsname\relax
  \def\citenamefont#1{#1}\fi
\expandafter\ifx\csname url\endcsname\relax
  \def\url#1{\texttt{#1}}\fi
\expandafter\ifx\csname urlprefix\endcsname\relax\def\urlprefix{URL }\fi
\providecommand{\bibinfo}[2]{#2}
\providecommand{\eprint}[2][]{\url{#2}}


\bibitem [{\citenamefont {Kirkpatrick}\ \emph {et~al.}(1983)\citenamefont
  {Kirkpatrick}, \citenamefont {Gelatt},\ and\ \citenamefont
  {Vecchi}}]{Kirkpatrick1983}%
  \BibitemOpen
  \bibfield  {author} {\bibinfo {author} {\bibfnamefont {S.}~\bibnamefont
  {Kirkpatrick}}, \bibinfo {author} {\bibfnamefont {C.~D.}\ \bibnamefont
  {Gelatt}},\ and\ \bibinfo {author} {\bibfnamefont {M.~P.}\ \bibnamefont
  {Vecchi}},\ }\href {https://doi.org/10.1126/science.220.4598.671} {\bibfield
  {journal} {\bibinfo  {journal} {Science}\ }\textbf {\bibinfo {volume}
  {220}},\ \bibinfo {pages} {671} (\bibinfo {year} {1983})},\ \eprint
  {https://www.science.org/doi/pdf/10.1126/science.220.4598.671} \BibitemShut
  {NoStop}%
\bibitem [{\citenamefont {Finnila}\ \emph {et~al.}(1994)\citenamefont
  {Finnila}, \citenamefont {Gomez}, \citenamefont {Sebenik}, \citenamefont
  {Stenson},\ and\ \citenamefont {Doll}}]{Finnila1994}%
  \BibitemOpen
  \bibfield  {author} {\bibinfo {author} {\bibfnamefont {A.}~\bibnamefont
  {Finnila}}, \bibinfo {author} {\bibfnamefont {M.}~\bibnamefont {Gomez}},
  \bibinfo {author} {\bibfnamefont {C.}~\bibnamefont {Sebenik}}, \bibinfo
  {author} {\bibfnamefont {C.}~\bibnamefont {Stenson}},\ and\ \bibinfo {author}
  {\bibfnamefont {J.}~\bibnamefont {Doll}},\ }\href
  {https://doi.org/https://doi.org/10.1016/0009-2614(94)00117-0} {\bibfield
  {journal} {\bibinfo  {journal} {Chemical Physics Letters}\ }\textbf {\bibinfo
  {volume} {219}},\ \bibinfo {pages} {343} (\bibinfo {year}
  {1994})}\BibitemShut {NoStop}%
\bibitem [{\citenamefont {Earl}\ and\ \citenamefont {Deem}(2005)}]{Earl2005}%
  \BibitemOpen
  \bibfield  {author} {\bibinfo {author} {\bibfnamefont {D.~J.}\ \bibnamefont
  {Earl}}\ and\ \bibinfo {author} {\bibfnamefont {M.~W.}\ \bibnamefont
  {Deem}},\ }\href {https://doi.org/10.1039/B509983H} {\bibfield  {journal}
  {\bibinfo  {journal} {Phys. Chem. Chem. Phys.}\ }\textbf {\bibinfo {volume}
  {7}},\ \bibinfo {pages} {3910} (\bibinfo {year} {2005})}\BibitemShut
  {NoStop}%
\bibitem [{\citenamefont {Zhang}\ \emph {et~al.}(2021)\citenamefont {Zhang},
  \citenamefont {Bengio}, \citenamefont {Hardt}, \citenamefont {Recht},\ and\
  \citenamefont {Vinyals}}]{Zhang2021}%
  \BibitemOpen
  \bibfield  {author} {\bibinfo {author} {\bibfnamefont {C.}~\bibnamefont
  {Zhang}}, \bibinfo {author} {\bibfnamefont {S.}~\bibnamefont {Bengio}},
  \bibinfo {author} {\bibfnamefont {M.}~\bibnamefont {Hardt}}, \bibinfo
  {author} {\bibfnamefont {B.}~\bibnamefont {Recht}},\ and\ \bibinfo {author}
  {\bibfnamefont {O.}~\bibnamefont {Vinyals}},\ }\href
  {https://doi.org/10.1145/3446776} {\bibfield  {journal} {\bibinfo  {journal}
  {Commun. ACM}\ }\textbf {\bibinfo {volume} {64}},\ \bibinfo {pages}
  {107–115} (\bibinfo {year} {2021})}\BibitemShut {NoStop}%
\bibitem [{\citenamefont {Choromanska}\ \emph {et~al.}(2015)\citenamefont
  {Choromanska}, \citenamefont {Henaff}, \citenamefont {Mathieu}, \citenamefont
  {Ben~Arous},\ and\ \citenamefont {LeCun}}]{Choromanska2015}%
  \BibitemOpen
  \bibfield  {author} {\bibinfo {author} {\bibfnamefont {A.}~\bibnamefont
  {Choromanska}}, \bibinfo {author} {\bibfnamefont {M.}~\bibnamefont {Henaff}},
  \bibinfo {author} {\bibfnamefont {M.}~\bibnamefont {Mathieu}}, \bibinfo
  {author} {\bibfnamefont {G.}~\bibnamefont {Ben~Arous}},\ and\ \bibinfo
  {author} {\bibfnamefont {Y.}~\bibnamefont {LeCun}},\ }in\ \href
  {https://proceedings.mlr.press/v38/choromanska15.html} {\emph {\bibinfo
  {booktitle} {Proceedings of the Eighteenth International Conference on
  Artificial Intelligence and Statistics}}},\ \bibinfo {series} {Proceedings of
  Machine Learning Research}, Vol.~\bibinfo {volume} {38},\ \bibinfo {editor}
  {edited by\ \bibinfo {editor} {\bibfnamefont {G.}~\bibnamefont {Lebanon}}\
  and\ \bibinfo {editor} {\bibfnamefont {S.~V.~N.}\ \bibnamefont
  {Vishwanathan}}}\ (\bibinfo  {publisher} {PMLR},\ \bibinfo {address} {San
  Diego, California, USA},\ \bibinfo {year} {2015})\ pp.\ \bibinfo {pages}
  {192--204}\BibitemShut {NoStop}%
\bibitem [{\citenamefont {Chaudhari}\ \emph {et~al.}(2019)\citenamefont
  {Chaudhari}, \citenamefont {Choromanska}, \citenamefont {Soatto},
  \citenamefont {LeCun}, \citenamefont {Baldassi}, \citenamefont {Borgs},
  \citenamefont {Chayes}, \citenamefont {Sagun},\ and\ \citenamefont
  {Zecchina}}]{Chaudhari2019}%
  \BibitemOpen
  \bibfield  {author} {\bibinfo {author} {\bibfnamefont {P.}~\bibnamefont
  {Chaudhari}}, \bibinfo {author} {\bibfnamefont {A.}~\bibnamefont
  {Choromanska}}, \bibinfo {author} {\bibfnamefont {S.}~\bibnamefont {Soatto}},
  \bibinfo {author} {\bibfnamefont {Y.}~\bibnamefont {LeCun}}, \bibinfo
  {author} {\bibfnamefont {C.}~\bibnamefont {Baldassi}}, \bibinfo {author}
  {\bibfnamefont {C.}~\bibnamefont {Borgs}}, \bibinfo {author} {\bibfnamefont
  {J.}~\bibnamefont {Chayes}}, \bibinfo {author} {\bibfnamefont
  {L.}~\bibnamefont {Sagun}},\ and\ \bibinfo {author} {\bibfnamefont
  {R.}~\bibnamefont {Zecchina}},\ }\href
  {https://doi.org/10.1088/1742-5468/ab39d9} {\bibfield  {journal} {\bibinfo
  {journal} {Journal of Statistical Mechanics: Theory and Experiment}\ }\textbf
  {\bibinfo {volume} {2019}},\ \bibinfo {pages} {124018} (\bibinfo {year}
  {2019})}\BibitemShut {NoStop}%
\bibitem [{\citenamefont {Baldassi}\ \emph {et~al.}(2016)\citenamefont
  {Baldassi}, \citenamefont {Borgs}, \citenamefont {Chayes}, \citenamefont
  {Ingrosso}, \citenamefont {Lucibello}, \citenamefont {Saglietti},\ and\
  \citenamefont {Zecchina}}]{Baldassi2016}%
  \BibitemOpen
  \bibfield  {author} {\bibinfo {author} {\bibfnamefont {C.}~\bibnamefont
  {Baldassi}}, \bibinfo {author} {\bibfnamefont {C.}~\bibnamefont {Borgs}},
  \bibinfo {author} {\bibfnamefont {J.~T.}\ \bibnamefont {Chayes}}, \bibinfo
  {author} {\bibfnamefont {A.}~\bibnamefont {Ingrosso}}, \bibinfo {author}
  {\bibfnamefont {C.}~\bibnamefont {Lucibello}}, \bibinfo {author}
  {\bibfnamefont {L.}~\bibnamefont {Saglietti}},\ and\ \bibinfo {author}
  {\bibfnamefont {R.}~\bibnamefont {Zecchina}},\ }\href
  {https://doi.org/10.1073/pnas.1608103113} {\bibfield  {journal} {\bibinfo
  {journal} {Proceedings of the National Academy of Sciences}\ }\textbf
  {\bibinfo {volume} {113}},\ \bibinfo {pages} {E7655} (\bibinfo {year}
  {2016})},\ \eprint
  {https://www.pnas.org/doi/pdf/10.1073/pnas.1608103113} \BibitemShut {NoStop}%
\bibitem [{\citenamefont {Baity-Jesi}\ \emph {et~al.}(2018)\citenamefont
  {Baity-Jesi}, \citenamefont {Calore}, \citenamefont {Cruz}, \citenamefont
  {Fernandez}, \citenamefont {Gil-Narvion}, \citenamefont {Gordillo-Guerrero},
  \citenamefont {I\~niguez}, \citenamefont {Maiorano}, \citenamefont
  {Marinari}, \citenamefont {Martin-Mayor}, \citenamefont {Moreno-Gordo},
  \citenamefont {Mu\~noz Sudupe}, \citenamefont {Navarro}, \citenamefont
  {Parisi}, \citenamefont {Perez-Gaviro}, \citenamefont {Ricci-Tersenghi},
  \citenamefont {Ruiz-Lorenzo}, \citenamefont {Schifano}, \citenamefont
  {Seoane}, \citenamefont {Tarancon}, \citenamefont {Tripiccione},\ and\
  \citenamefont {Yllanes}}]{Baity-Jesi2018}%
  \BibitemOpen
  \bibfield  {author} {\bibinfo {author} {\bibfnamefont {M.}~\bibnamefont
  {Baity-Jesi}}, \bibinfo {author} {\bibfnamefont {E.}~\bibnamefont {Calore}},
  \bibinfo {author} {\bibfnamefont {A.}~\bibnamefont {Cruz}}, \bibinfo {author}
  {\bibfnamefont {L.~A.}\ \bibnamefont {Fernandez}}, \bibinfo {author}
  {\bibfnamefont {J.~M.}\ \bibnamefont {Gil-Narvion}}, \bibinfo {author}
  {\bibfnamefont {A.}~\bibnamefont {Gordillo-Guerrero}}, \bibinfo {author}
  {\bibfnamefont {D.}~\bibnamefont {I\~niguez}}, \bibinfo {author}
  {\bibfnamefont {A.}~\bibnamefont {Maiorano}}, \bibinfo {author}
  {\bibfnamefont {E.}~\bibnamefont {Marinari}}, \bibinfo {author}
  {\bibfnamefont {V.}~\bibnamefont {Martin-Mayor}}, \bibinfo {author}
  {\bibfnamefont {J.}~\bibnamefont {Moreno-Gordo}}, \bibinfo {author}
  {\bibfnamefont {A.}~\bibnamefont {Mu\~noz Sudupe}}, \bibinfo {author}
  {\bibfnamefont {D.}~\bibnamefont {Navarro}}, \bibinfo {author} {\bibfnamefont
  {G.}~\bibnamefont {Parisi}}, \bibinfo {author} {\bibfnamefont
  {S.}~\bibnamefont {Perez-Gaviro}}, \bibinfo {author} {\bibfnamefont
  {F.}~\bibnamefont {Ricci-Tersenghi}}, \bibinfo {author} {\bibfnamefont
  {J.~J.}\ \bibnamefont {Ruiz-Lorenzo}}, \bibinfo {author} {\bibfnamefont
  {S.~F.}\ \bibnamefont {Schifano}}, \bibinfo {author} {\bibfnamefont
  {B.}~\bibnamefont {Seoane}}, \bibinfo {author} {\bibfnamefont
  {A.}~\bibnamefont {Tarancon}}, \bibinfo {author} {\bibfnamefont
  {R.}~\bibnamefont {Tripiccione}},\ and\ \bibinfo {author} {\bibfnamefont
  {D.}~\bibnamefont {Yllanes}} (\bibinfo {collaboration} {Janus
  Collaboration}),\ }\href {https://doi.org/10.1103/PhysRevLett.120.267203}
  {\bibfield  {journal} {\bibinfo  {journal} {Phys. Rev. Lett.}\ }\textbf
  {\bibinfo {volume} {120}},\ \bibinfo {pages} {267203} (\bibinfo {year}
  {2018})}\BibitemShut {NoStop}%
\bibitem [{\citenamefont {Hartnett}\ \emph {et~al.}(2018)\citenamefont
  {Hartnett}, \citenamefont {Parker},\ and\ \citenamefont
  {Geist}}]{Hartnett2018}%
  \BibitemOpen
  \bibfield  {author} {\bibinfo {author} {\bibfnamefont {G.~S.}\ \bibnamefont
  {Hartnett}}, \bibinfo {author} {\bibfnamefont {E.}~\bibnamefont {Parker}},\
  and\ \bibinfo {author} {\bibfnamefont {E.}~\bibnamefont {Geist}},\ }\href
  {https://doi.org/10.1103/PhysRevE.98.022116} {\bibfield  {journal} {\bibinfo
  {journal} {Phys. Rev. E}\ }\textbf {\bibinfo {volume} {98}},\ \bibinfo
  {pages} {022116} (\bibinfo {year} {2018})}\BibitemShut {NoStop}%
\bibitem [{\citenamefont {Sarao~Mannelli}\ \emph {et~al.}(2020)\citenamefont
  {Sarao~Mannelli}, \citenamefont {Biroli}, \citenamefont {Cammarota},
  \citenamefont {Krzakala}, \citenamefont {Urbani},\ and\ \citenamefont
  {Zdeborov\'a}}]{Mannelli2020}%
  \BibitemOpen
  \bibfield  {author} {\bibinfo {author} {\bibfnamefont {S.}~\bibnamefont
  {Sarao~Mannelli}}, \bibinfo {author} {\bibfnamefont {G.}~\bibnamefont
  {Biroli}}, \bibinfo {author} {\bibfnamefont {C.}~\bibnamefont {Cammarota}},
  \bibinfo {author} {\bibfnamefont {F.}~\bibnamefont {Krzakala}}, \bibinfo
  {author} {\bibfnamefont {P.}~\bibnamefont {Urbani}},\ and\ \bibinfo {author}
  {\bibfnamefont {L.}~\bibnamefont {Zdeborov\'a}},\ }\href
  {https://doi.org/10.1103/PhysRevX.10.011057} {\bibfield  {journal} {\bibinfo
  {journal} {Phys. Rev. X}\ }\textbf {\bibinfo {volume} {10}},\ \bibinfo
  {pages} {011057} (\bibinfo {year} {2020})}\BibitemShut {NoStop}%
\bibitem [{\citenamefont {Mignacco}\ and\ \citenamefont
  {Urbani}(2022)}]{Mignacco2022}%
  \BibitemOpen
  \bibfield  {author} {\bibinfo {author} {\bibfnamefont {F.}~\bibnamefont
  {Mignacco}}\ and\ \bibinfo {author} {\bibfnamefont {P.}~\bibnamefont
  {Urbani}},\ }\href {https://doi.org/10.1088/1742-5468/ac841d} {\bibfield
  {journal} {\bibinfo  {journal} {Journal of Statistical Mechanics: Theory and
  Experiment}\ }\textbf {\bibinfo {volume} {2022}},\ \bibinfo {pages} {083405}
  (\bibinfo {year} {2022})}\BibitemShut {NoStop}%
\bibitem [{\citenamefont {Manacorda}\ and\ \citenamefont
  {Zamponi}(2022)}]{Manacorda2022}%
  \BibitemOpen
  \bibfield  {author} {\bibinfo {author} {\bibfnamefont {A.}~\bibnamefont
  {Manacorda}}\ and\ \bibinfo {author} {\bibfnamefont {F.}~\bibnamefont
  {Zamponi}},\ }\href {https://doi.org/10.1088/1751-8121/ac7f06} {\bibfield
  {journal} {\bibinfo  {journal} {Journal of Physics A: Mathematical and
  Theoretical}\ }\textbf {\bibinfo {volume} {55}},\ \bibinfo {pages} {334001}
  (\bibinfo {year} {2022})}\BibitemShut {NoStop}%
\bibitem [{\citenamefont {Geiger}\ \emph {et~al.}(2019)\citenamefont {Geiger},
  \citenamefont {Spigler}, \citenamefont {d'Ascoli}, \citenamefont {Sagun},
  \citenamefont {Baity-Jesi}, \citenamefont {Biroli},\ and\ \citenamefont
  {Wyart}}]{Geiger2019}%
  \BibitemOpen
  \bibfield  {author} {\bibinfo {author} {\bibfnamefont {M.}~\bibnamefont
  {Geiger}}, \bibinfo {author} {\bibfnamefont {S.}~\bibnamefont {Spigler}},
  \bibinfo {author} {\bibfnamefont {S.}~\bibnamefont {d'Ascoli}}, \bibinfo
  {author} {\bibfnamefont {L.}~\bibnamefont {Sagun}}, \bibinfo {author}
  {\bibfnamefont {M.}~\bibnamefont {Baity-Jesi}}, \bibinfo {author}
  {\bibfnamefont {G.}~\bibnamefont {Biroli}},\ and\ \bibinfo {author}
  {\bibfnamefont {M.}~\bibnamefont {Wyart}},\ }\href
  {https://doi.org/10.1103/PhysRevE.100.012115} {\bibfield  {journal} {\bibinfo
   {journal} {Phys. Rev. E}\ }\textbf {\bibinfo {volume} {100}},\ \bibinfo
  {pages} {012115} (\bibinfo {year} {2019})}\BibitemShut {NoStop}%
\bibitem [{\citenamefont {Cugliandolo}\ and\ \citenamefont
  {Kurchan}(1993)}]{Cugliandolo1993}%
  \BibitemOpen
  \bibfield  {author} {\bibinfo {author} {\bibfnamefont {L.~F.}\ \bibnamefont
  {Cugliandolo}}\ and\ \bibinfo {author} {\bibfnamefont {J.}~\bibnamefont
  {Kurchan}},\ }\href {https://doi.org/10.1103/PhysRevLett.71.173} {\bibfield
  {journal} {\bibinfo  {journal} {Phys. Rev. Lett.}\ }\textbf {\bibinfo
  {volume} {71}},\ \bibinfo {pages} {173} (\bibinfo {year} {1993})}\BibitemShut
  {NoStop}%
\bibitem [{\citenamefont {Cugliandolo}\ and\ \citenamefont
  {Kurchan}(1994)}]{Cugliandolo1994}%
  \BibitemOpen
  \bibfield  {author} {\bibinfo {author} {\bibfnamefont {L.~F.}\ \bibnamefont
  {Cugliandolo}}\ and\ \bibinfo {author} {\bibfnamefont {J.}~\bibnamefont
  {Kurchan}},\ }\href {https://doi.org/10.1088/0305-4470/27/17/011} {\bibfield
  {journal} {\bibinfo  {journal} {Journal of Physics A: Mathematical and
  General}\ }\textbf {\bibinfo {volume} {27}},\ \bibinfo {pages} {5749}
  (\bibinfo {year} {1994})}\BibitemShut {NoStop}%
\bibitem [{\citenamefont {De~Dominicis}\ and\ \citenamefont
  {Kondor}(1983)}]{DeDominicis1983}%
  \BibitemOpen
  \bibfield  {author} {\bibinfo {author} {\bibfnamefont {C.}~\bibnamefont
  {De~Dominicis}}\ and\ \bibinfo {author} {\bibfnamefont {I.}~\bibnamefont
  {Kondor}},\ }\href {https://doi.org/10.1103/PhysRevB.27.606} {\bibfield
  {journal} {\bibinfo  {journal} {Phys. Rev. B}\ }\textbf {\bibinfo {volume}
  {27}},\ \bibinfo {pages} {606} (\bibinfo {year} {1983})}\BibitemShut
  {NoStop}%
\bibitem [{\citenamefont {P\'azm\'andi}\ \emph {et~al.}(1999)\citenamefont
  {P\'azm\'andi}, \citenamefont {Zar\'and},\ and\ \citenamefont
  {Zim\'anyi}}]{Pazmandi1999}%
  \BibitemOpen
  \bibfield  {author} {\bibinfo {author} {\bibfnamefont {F.}~\bibnamefont
  {P\'azm\'andi}}, \bibinfo {author} {\bibfnamefont {G.}~\bibnamefont
  {Zar\'and}},\ and\ \bibinfo {author} {\bibfnamefont {G.~T.}\ \bibnamefont
  {Zim\'anyi}},\ }\href {https://doi.org/10.1103/PhysRevLett.83.1034}
  {\bibfield  {journal} {\bibinfo  {journal} {Phys. Rev. Lett.}\ }\textbf
  {\bibinfo {volume} {83}},\ \bibinfo {pages} {1034} (\bibinfo {year}
  {1999})}\BibitemShut {NoStop}%
\bibitem [{\citenamefont {Le~Doussal}\ \emph {et~al.}(2010)\citenamefont
  {Le~Doussal}, \citenamefont {Müller},\ and\ \citenamefont
  {Wiese}}]{LeDoussal2010}%
  \BibitemOpen
  \bibfield  {author} {\bibinfo {author} {\bibfnamefont {P.}~\bibnamefont
  {Le~Doussal}}, \bibinfo {author} {\bibfnamefont {M.}~\bibnamefont
  {Müller}},\ and\ \bibinfo {author} {\bibfnamefont {K.~J.}\ \bibnamefont
  {Wiese}},\ }\href {https://doi.org/10.1209/0295-5075/91/57004} {\bibfield
  {journal} {\bibinfo  {journal} {Europhysics Letters}\ }\textbf {\bibinfo
  {volume} {91}},\ \bibinfo {pages} {57004} (\bibinfo {year}
  {2010})}\BibitemShut {NoStop}%
\bibitem [{\citenamefont {Cugliandolo}\ and\ \citenamefont
  {Kurchan}(1995)}]{Cugliandolo1995}%
  \BibitemOpen
  \bibfield  {author} {\bibinfo {author} {\bibfnamefont {L.~F.}\ \bibnamefont
  {Cugliandolo}}\ and\ \bibinfo {author} {\bibfnamefont {J.}~\bibnamefont
  {Kurchan}},\ }\href {https://doi.org/10.1080/01418639508238541} {\bibfield
  {journal} {\bibinfo  {journal} {Philosophical Magazine B}\ }\textbf {\bibinfo
  {volume} {71}},\ \bibinfo {pages} {501} (\bibinfo {year} {1995})}\BibitemShut
  {NoStop}%
\bibitem [{\citenamefont {Kadanoff}\ and\ \citenamefont
  {Baym}(1962)}]{Kadanoff_book}%
  \BibitemOpen
  \bibfield  {author} {\bibinfo {author} {\bibfnamefont {L.}~\bibnamefont
  {Kadanoff}}\ and\ \bibinfo {author} {\bibfnamefont {G.}~\bibnamefont
  {Baym}},\ }\href {https://books.google.de/books?id=1-FEAAAAIAAJ} {\emph
  {\bibinfo {title} {{Q}uantum statistical mechanics: {G}reen's function
  methods in equilibrium and nonequilibrium problems}}},\ Frontiers in physics\
  (\bibinfo  {publisher} {W.A. Benjamin},\ \bibinfo {year} {1962})\BibitemShut
  {NoStop}%
\bibitem [{\citenamefont {Baym}(1962)}]{Baym1962}%
  \BibitemOpen
  \bibfield  {author} {\bibinfo {author} {\bibfnamefont {G.}~\bibnamefont
  {Baym}},\ }\href {https://doi.org/10.1103/PhysRev.127.1391} {\bibfield
  {journal} {\bibinfo  {journal} {Phys. Rev.}\ }\textbf {\bibinfo {volume}
  {127}},\ \bibinfo {pages} {1391} (\bibinfo {year} {1962})}\BibitemShut
  {NoStop}%
\bibitem [{\citenamefont {Cornwall}\ \emph {et~al.}(1974)\citenamefont
  {Cornwall}, \citenamefont {Jackiw},\ and\ \citenamefont
  {Tomboulis}}]{Cornwall1974}%
  \BibitemOpen
  \bibfield  {author} {\bibinfo {author} {\bibfnamefont {J.~M.}\ \bibnamefont
  {Cornwall}}, \bibinfo {author} {\bibfnamefont {R.}~\bibnamefont {Jackiw}},\
  and\ \bibinfo {author} {\bibfnamefont {E.}~\bibnamefont {Tomboulis}},\ }\href
  {https://doi.org/10.1103/PhysRevD.10.2428} {\bibfield  {journal} {\bibinfo
  {journal} {Phys. Rev. D}\ }\textbf {\bibinfo {volume} {10}},\ \bibinfo
  {pages} {2428} (\bibinfo {year} {1974})}\BibitemShut {NoStop}%
\bibitem [{\citenamefont {Fetter}\ and\ \citenamefont
  {Walecka}(2003)}]{Fetter2003}%
  \BibitemOpen
  \bibfield  {author} {\bibinfo {author} {\bibfnamefont {A.}~\bibnamefont
  {Fetter}}\ and\ \bibinfo {author} {\bibfnamefont {J.}~\bibnamefont
  {Walecka}},\ }\href {https://books.google.de/books?id=0wekf1s83b0C} {\emph
  {\bibinfo {title} {Quantum Theory of Many-particle Systems}}},\ Dover Books
  on Physics\ (\bibinfo  {publisher} {Dover Publications},\ \bibinfo {year}
  {2003})\BibitemShut {NoStop}%
\bibitem [{\citenamefont {Altland}\ and\ \citenamefont
  {Simons}(2010)}]{Altland2010}%
  \BibitemOpen
  \bibfield  {author} {\bibinfo {author} {\bibfnamefont {A.}~\bibnamefont
  {Altland}}\ and\ \bibinfo {author} {\bibfnamefont {B.}~\bibnamefont
  {Simons}},\ }\href {https://books.google.de/books?id=GpF0Pgo8CqAC} {\emph
  {\bibinfo {title} {Condensed Matter Field Theory}}},\ Cambridge books online\
  (\bibinfo  {publisher} {Cambridge University Press},\ \bibinfo {year}
  {2010})\BibitemShut {NoStop}%
\bibitem [{\citenamefont {Berges}\ \emph {et~al.}(2008)\citenamefont {Berges},
  \citenamefont {Rothkopf},\ and\ \citenamefont {Schmidt}}]{Berges2008}%
  \BibitemOpen
  \bibfield  {author} {\bibinfo {author} {\bibfnamefont {J.}~\bibnamefont
  {Berges}}, \bibinfo {author} {\bibfnamefont {A.}~\bibnamefont {Rothkopf}},\
  and\ \bibinfo {author} {\bibfnamefont {J.}~\bibnamefont {Schmidt}},\ }\href
  {https://doi.org/10.1103/PhysRevLett.101.041603} {\bibfield  {journal}
  {\bibinfo  {journal} {Phys. Rev. Lett.}\ }\textbf {\bibinfo {volume} {101}},\
  \bibinfo {pages} {041603} (\bibinfo {year} {2008})}\BibitemShut {NoStop}%
\bibitem [{\citenamefont {Folena}\ \emph {et~al.}(2020)\citenamefont {Folena},
  \citenamefont {Franz},\ and\ \citenamefont {Ricci-Tersenghi}}]{Folena2020}%
  \BibitemOpen
  \bibfield  {author} {\bibinfo {author} {\bibfnamefont {G.}~\bibnamefont
  {Folena}}, \bibinfo {author} {\bibfnamefont {S.}~\bibnamefont {Franz}},\ and\
  \bibinfo {author} {\bibfnamefont {F.}~\bibnamefont {Ricci-Tersenghi}},\
  }\href {https://doi.org/10.1103/PhysRevX.10.031045} {\bibfield  {journal}
  {\bibinfo  {journal} {Phys. Rev. X}\ }\textbf {\bibinfo {volume} {10}},\
  \bibinfo {pages} {031045} (\bibinfo {year} {2020})}\BibitemShut {NoStop}%
\bibitem [{\citenamefont {Köhler}\ \emph {et~al.}(1999)\citenamefont
  {Köhler}, \citenamefont {Kwong},\ and\ \citenamefont
  {Yousif}}]{Koehler1999}%
  \BibitemOpen
  \bibfield  {author} {\bibinfo {author} {\bibfnamefont {H.}~\bibnamefont
  {Köhler}}, \bibinfo {author} {\bibfnamefont {N.}~\bibnamefont {Kwong}},\
  and\ \bibinfo {author} {\bibfnamefont {H.~A.}\ \bibnamefont {Yousif}},\
  }\href {https://doi.org/https://doi.org/10.1016/S0010-4655(99)00260-X}
  {\bibfield  {journal} {\bibinfo  {journal} {Computer Physics Communications}\
  }\textbf {\bibinfo {volume} {123}},\ \bibinfo {pages} {123} (\bibinfo {year}
  {1999})}\BibitemShut {NoStop}%
\bibitem [{\citenamefont {Aoki}\ \emph {et~al.}(2014)\citenamefont {Aoki},
  \citenamefont {Tsuji}, \citenamefont {Eckstein}, \citenamefont {Kollar},
  \citenamefont {Oka},\ and\ \citenamefont {Werner}}]{Aoki2014}%
  \BibitemOpen
  \bibfield  {author} {\bibinfo {author} {\bibfnamefont {H.}~\bibnamefont
  {Aoki}}, \bibinfo {author} {\bibfnamefont {N.}~\bibnamefont {Tsuji}},
  \bibinfo {author} {\bibfnamefont {M.}~\bibnamefont {Eckstein}}, \bibinfo
  {author} {\bibfnamefont {M.}~\bibnamefont {Kollar}}, \bibinfo {author}
  {\bibfnamefont {T.}~\bibnamefont {Oka}},\ and\ \bibinfo {author}
  {\bibfnamefont {P.}~\bibnamefont {Werner}},\ }\href
  {https://doi.org/10.1103/RevModPhys.86.779} {\bibfield  {journal} {\bibinfo
  {journal} {Rev. Mod. Phys.}\ }\textbf {\bibinfo {volume} {86}},\ \bibinfo
  {pages} {779} (\bibinfo {year} {2014})}\BibitemShut {NoStop}%
\bibitem [{\citenamefont {{J. P. Bouchaud}}(1992)}]{Bouchaud1992}%
  \BibitemOpen
  \bibfield  {author} {\bibinfo {author} {\bibnamefont {{J. P. Bouchaud}}},\
  }\href {https://doi.org/10.1051/jp1:1992238} {\bibfield  {journal} {\bibinfo
  {journal} {J. Phys. I France}\ }\textbf {\bibinfo {volume} {2}},\ \bibinfo
  {pages} {1705} (\bibinfo {year} {1992})}\BibitemShut {NoStop}%
\bibitem [{\citenamefont {Stahl}\ \emph {et~al.}(2022)\citenamefont {Stahl},
  \citenamefont {Dasari}, \citenamefont {Li}, \citenamefont {Picano},
  \citenamefont {Werner},\ and\ \citenamefont {Eckstein}}]{Stahl2022}%
  \BibitemOpen
  \bibfield  {author} {\bibinfo {author} {\bibfnamefont {C.}~\bibnamefont
  {Stahl}}, \bibinfo {author} {\bibfnamefont {N.}~\bibnamefont {Dasari}},
  \bibinfo {author} {\bibfnamefont {J.}~\bibnamefont {Li}}, \bibinfo {author}
  {\bibfnamefont {A.}~\bibnamefont {Picano}}, \bibinfo {author} {\bibfnamefont
  {P.}~\bibnamefont {Werner}},\ and\ \bibinfo {author} {\bibfnamefont
  {M.}~\bibnamefont {Eckstein}},\ }\href
  {https://doi.org/10.1103/PhysRevB.105.115146} {\bibfield  {journal} {\bibinfo
   {journal} {Phys. Rev. B}\ }\textbf {\bibinfo {volume} {105}},\ \bibinfo
  {pages} {115146} (\bibinfo {year} {2022})}\BibitemShut {NoStop}%
\bibitem [{\citenamefont {Lipavsk\'y}\ \emph {et~al.}(1986)\citenamefont
  {Lipavsk\'y}, \citenamefont {\ifmmode \check{S}\else
  \v{S}\fi{}pi\ifmmode~\check{c}\else \v{c}\fi{}ka},\ and\ \citenamefont
  {Velick\'y}}]{Lipavsky1986}%
  \BibitemOpen
  \bibfield  {author} {\bibinfo {author} {\bibfnamefont {P.}~\bibnamefont
  {Lipavsk\'y}}, \bibinfo {author} {\bibfnamefont {V.}~\bibnamefont {\ifmmode
  \check{S}\else \v{S}\fi{}pi\ifmmode~\check{c}\else \v{c}\fi{}ka}},\ and\
  \bibinfo {author} {\bibfnamefont {B.}~\bibnamefont {Velick\'y}},\ }\href
  {https://doi.org/10.1103/PhysRevB.34.6933} {\bibfield  {journal} {\bibinfo
  {journal} {Phys. Rev. B}\ }\textbf {\bibinfo {volume} {34}},\ \bibinfo
  {pages} {6933} (\bibinfo {year} {1986})}\BibitemShut {NoStop}%
\bibitem [{\citenamefont {Kamenev}(2011)}]{Kamenev_book}%
  \BibitemOpen
  \bibfield  {author} {\bibinfo {author} {\bibfnamefont {A.}~\bibnamefont
  {Kamenev}},\ }\href {http://books.google.de/books?id=CwlrUepnla4C} {\emph
  {\bibinfo {title} {{F}ield {T}heory of {N}on-{E}quilibrium {S}ystems}}}\
  (\bibinfo  {publisher} {Cambridge University Press},\ \bibinfo {year}
  {2011})\BibitemShut {NoStop}%
\bibitem [{\citenamefont {Ketcheson}(2008)}]{Ketcheson2008}%
  \BibitemOpen
  \bibfield  {author} {\bibinfo {author} {\bibfnamefont {D.~I.}\ \bibnamefont
  {Ketcheson}},\ }\href {https://doi.org/10.1137/07070485X} {\bibfield
  {journal} {\bibinfo  {journal} {SIAM Journal on Scientific Computing}\
  }\textbf {\bibinfo {volume} {30}},\ \bibinfo {pages} {2113} (\bibinfo {year}
  {2008})},\ \eprint
  {https://doi.org/10.1137/07070485X} \BibitemShut {NoStop}%
\bibitem [{\citenamefont {Fekete}\ \emph {et~al.}(2022)\citenamefont {Fekete},
  \citenamefont {Conde},\ and\ \citenamefont {Shadid}}]{Fekete2022}%
  \BibitemOpen
  \bibfield  {author} {\bibinfo {author} {\bibfnamefont {I.}~\bibnamefont
  {Fekete}}, \bibinfo {author} {\bibfnamefont {S.}~\bibnamefont {Conde}},\ and\
  \bibinfo {author} {\bibfnamefont {J.~N.}\ \bibnamefont {Shadid}},\ }\href
  {https://doi.org/https://doi.org/10.1016/j.cam.2022.114325} {\bibfield
  {journal} {\bibinfo  {journal} {Journal of Computational and Applied
  Mathematics}\ }\textbf {\bibinfo {volume} {412}},\ \bibinfo {pages} {114325}
  (\bibinfo {year} {2022})}\BibitemShut {NoStop}%
\bibitem [{\citenamefont {Kim}\ and\ \citenamefont {Latz}(2001)}]{Kim2001}%
  \BibitemOpen
  \bibfield  {author} {\bibinfo {author} {\bibfnamefont {B.}~\bibnamefont
  {Kim}}\ and\ \bibinfo {author} {\bibfnamefont {A.}~\bibnamefont {Latz}},\
  }\href {https://doi.org/10.1209/epl/i2001-00202-4} {\bibfield  {journal}
  {\bibinfo  {journal} {Europhysics Letters}\ }\textbf {\bibinfo {volume}
  {53}},\ \bibinfo {pages} {660} (\bibinfo {year} {2001})}\BibitemShut
  {NoStop}%
\bibitem [{\citenamefont {Berthier}\ \emph {et~al.}(2007)\citenamefont
  {Berthier}, \citenamefont {Biroli}, \citenamefont {Bouchaud}, \citenamefont
  {Kob}, \citenamefont {Miyazaki},\ and\ \citenamefont
  {Reichman}}]{Berthier2007}%
  \BibitemOpen
  \bibfield  {author} {\bibinfo {author} {\bibfnamefont {L.}~\bibnamefont
  {Berthier}}, \bibinfo {author} {\bibfnamefont {G.}~\bibnamefont {Biroli}},
  \bibinfo {author} {\bibfnamefont {J.-P.}\ \bibnamefont {Bouchaud}}, \bibinfo
  {author} {\bibfnamefont {W.}~\bibnamefont {Kob}}, \bibinfo {author}
  {\bibfnamefont {K.}~\bibnamefont {Miyazaki}},\ and\ \bibinfo {author}
  {\bibfnamefont {D.~R.}\ \bibnamefont {Reichman}},\ }\href
  {https://doi.org/10.1063/1.2721555} {\bibfield  {journal} {\bibinfo
  {journal} {The Journal of Chemical Physics}\ }\textbf {\bibinfo {volume}
  {126}},\ \bibinfo {pages} {184504} (\bibinfo {year} {2007})},\ \eprint
  {https://pubs.aip.org/aip/jcp/article-pdf/doi/10.1063/1.2721555/15398786/184504\_1\_online.pdf}
  \BibitemShut {NoStop}%
\bibitem [{\citenamefont {Folena}(2020)}]{Folena2020b}%
  \BibitemOpen
  \bibfield  {author} {\bibinfo {author} {\bibfnamefont {G.}~\bibnamefont
  {Folena}},\ }\emph {\bibinfo {title} {{The mixed p-spin model: selecting,
  following and losing states}}},\ \href
  {https://theses.hal.science/tel-02883385} {\bibinfo {type} {Theses}},\
  \bibinfo  {school} {{Universit{\'e} Paris-Saclay ; Universit{\`a} degli studi
  La Sapienza (Rome)}} (\bibinfo {year} {2020})\BibitemShut {NoStop}%
\bibitem [{\citenamefont {Barrat}\ \emph {et~al.}(1997)\citenamefont {Barrat},
  \citenamefont {Franz},\ and\ \citenamefont {Parisi}}]{Barrat1997}%
  \BibitemOpen
  \bibfield  {author} {\bibinfo {author} {\bibfnamefont {A.}~\bibnamefont
  {Barrat}}, \bibinfo {author} {\bibfnamefont {S.}~\bibnamefont {Franz}},\ and\
  \bibinfo {author} {\bibfnamefont {G.}~\bibnamefont {Parisi}},\ }\href
  {https://doi.org/10.1088/0305-4470/30/16/006} {\bibfield  {journal} {\bibinfo
   {journal} {Journal of Physics A: Mathematical and General}\ }\textbf
  {\bibinfo {volume} {30}},\ \bibinfo {pages} {5593} (\bibinfo {year}
  {1997})}\BibitemShut {NoStop}%
\bibitem [{\citenamefont {Crisanti}\ and\ \citenamefont
  {Leuzzi}(2004)}]{Crisanti2004}%
  \BibitemOpen
  \bibfield  {author} {\bibinfo {author} {\bibfnamefont {A.}~\bibnamefont
  {Crisanti}}\ and\ \bibinfo {author} {\bibfnamefont {L.}~\bibnamefont
  {Leuzzi}},\ }\href {https://doi.org/10.1103/PhysRevLett.93.217203} {\bibfield
   {journal} {\bibinfo  {journal} {Phys. Rev. Lett.}\ }\textbf {\bibinfo
  {volume} {93}},\ \bibinfo {pages} {217203} (\bibinfo {year}
  {2004})}\BibitemShut {NoStop}%
\bibitem [{\citenamefont {Crisanti}\ and\ \citenamefont
  {Leuzzi}(2006)}]{Crisanti2006}%
  \BibitemOpen
  \bibfield  {author} {\bibinfo {author} {\bibfnamefont {A.}~\bibnamefont
  {Crisanti}}\ and\ \bibinfo {author} {\bibfnamefont {L.}~\bibnamefont
  {Leuzzi}},\ }\href {https://doi.org/10.1103/PhysRevB.73.014412} {\bibfield
  {journal} {\bibinfo  {journal} {Phys. Rev. B}\ }\textbf {\bibinfo {volume}
  {73}},\ \bibinfo {pages} {014412} (\bibinfo {year} {2006})}\BibitemShut
  {NoStop}%
\bibitem [{\citenamefont {G\"otze}\ and\ \citenamefont
  {Sjogren}(1992)}]{Goetze1992}%
  \BibitemOpen
  \bibfield  {author} {\bibinfo {author} {\bibfnamefont {W.}~\bibnamefont
  {G\"otze}}\ and\ \bibinfo {author} {\bibfnamefont {L.}~\bibnamefont
  {Sjogren}},\ }\href {https://doi.org/10.1088/0034-4885/55/3/001} {\bibfield
  {journal} {\bibinfo  {journal} {Reports on Progress in Physics}\ }\textbf
  {\bibinfo {volume} {55}},\ \bibinfo {pages} {241} (\bibinfo {year}
  {1992})}\BibitemShut {NoStop}%
\bibitem [{\citenamefont {Monasson}\ and\ \citenamefont
  {Zecchina}(1997)}]{Monasson1997}%
  \BibitemOpen
  \bibfield  {author} {\bibinfo {author} {\bibfnamefont {R.}~\bibnamefont
  {Monasson}}\ and\ \bibinfo {author} {\bibfnamefont {R.}~\bibnamefont
  {Zecchina}},\ }\href {https://doi.org/10.1103/PhysRevE.56.1357} {\bibfield
  {journal} {\bibinfo  {journal} {Phys. Rev. E}\ }\textbf {\bibinfo {volume}
  {56}},\ \bibinfo {pages} {1357} (\bibinfo {year} {1997})}\BibitemShut
  {NoStop}%
\bibitem [{\citenamefont {{Silvio Franz}}\ and\ \citenamefont {{Giorgio
  Parisi}}(1995)}]{Franz1995b}%
  \BibitemOpen
  \bibfield  {author} {\bibinfo {author} {\bibnamefont {{Silvio Franz}}}\ and\
  \bibinfo {author} {\bibnamefont {{Giorgio Parisi}}},\ }\href
  {https://doi.org/10.1051/jp1:1995201} {\bibfield  {journal} {\bibinfo
  {journal} {J. Phys. I France}\ }\textbf {\bibinfo {volume} {5}},\ \bibinfo
  {pages} {1401} (\bibinfo {year} {1995})}\BibitemShut {NoStop}%
\bibitem [{\citenamefont {Ben~Arous}\ \emph {et~al.}(2006)\citenamefont
  {Ben~Arous}, \citenamefont {Dembo},\ and\ \citenamefont
  {Guionnet}}]{Arous2006}%
  \BibitemOpen
  \bibfield  {author} {\bibinfo {author} {\bibfnamefont {G.}~\bibnamefont
  {Ben~Arous}}, \bibinfo {author} {\bibfnamefont {A.}~\bibnamefont {Dembo}},\
  and\ \bibinfo {author} {\bibfnamefont {A.}~\bibnamefont {Guionnet}},\ }\href
  {https://doi.org/10.1007/s00440-005-0491-y} {\bibfield  {journal} {\bibinfo
  {journal} {Probability Theory and Related Fields}\ }\textbf {\bibinfo
  {volume} {136}},\ \bibinfo {pages} {619} (\bibinfo {year}
  {2006})}\BibitemShut {NoStop}%
\bibitem [{\citenamefont {Auffinger}\ and\ \citenamefont
  {Zhou}(2022)}]{Auffinger2022}%
  \BibitemOpen
  \bibfield  {author} {\bibinfo {author} {\bibfnamefont {A.}~\bibnamefont
  {Auffinger}}\ and\ \bibinfo {author} {\bibfnamefont {Y.}~\bibnamefont
  {Zhou}},\ }\href {https://arxiv.org/abs/2209.03866} {\bibinfo {title} {The
  spherical p+s spin glass at zero temperature}} (\bibinfo {year} {2022}),\
  \eprint {https://arxiv.org/abs/2209.03866} {arXiv:2209.03866 [math.PR]}
  \BibitemShut {NoStop}%
\bibitem [{\citenamefont {Cugliandolo}(2024)}]{Cugliandolo2023}%
  \BibitemOpen
  \bibfield  {author} {\bibinfo {author} {\bibfnamefont {L.~F.}\ \bibnamefont
  {Cugliandolo}},\ }\href
  {https://doi.org/https://doi.org/10.1146/annurev-conmatphys-040721-022848}
  {\bibfield  {journal} {\bibinfo  {journal} {Annual Review of Condensed Matter
  Physics}\ }\textbf {\bibinfo {volume} {15}},\ \bibinfo {pages} {177}
  (\bibinfo {year} {2024})}\BibitemShut {NoStop}%
\bibitem [{\citenamefont {Folena}\ and\ \citenamefont
  {Zamponi}(2023)}]{Folena2023}%
  \BibitemOpen
  \bibfield  {author} {\bibinfo {author} {\bibfnamefont {G.}~\bibnamefont
  {Folena}}\ and\ \bibinfo {author} {\bibfnamefont {F.}~\bibnamefont
  {Zamponi}},\ }\href {https://doi.org/10.21468/SciPostPhys.15.3.109}
  {\bibfield  {journal} {\bibinfo  {journal} {SciPost Phys.}\ }\textbf
  {\bibinfo {volume} {15}},\ \bibinfo {pages} {109} (\bibinfo {year}
  {2023})}\BibitemShut {NoStop}%
\bibitem [{\citenamefont {Götze}(2008)}]{Goetze2008}%
  \BibitemOpen
  \bibfield  {author} {\bibinfo {author} {\bibfnamefont {W.}~\bibnamefont
  {Götze}},\ }\href
  {https://doi.org/10.1093/acprof:oso/9780199235346.001.0001} {\emph {\bibinfo
  {title} {Complex Dynamics of Glass-Forming Liquids: A Mode-Coupling
  Theory}}}\ (\bibinfo  {publisher} {Oxford University Press},\ \bibinfo {year}
  {2008})\BibitemShut {NoStop}%
\bibitem [{\citenamefont {Cavagna}\ \emph {et~al.}(1998)\citenamefont
  {Cavagna}, \citenamefont {Giardina},\ and\ \citenamefont
  {Parisi}}]{Cavagna1998}%
  \BibitemOpen
  \bibfield  {author} {\bibinfo {author} {\bibfnamefont {A.}~\bibnamefont
  {Cavagna}}, \bibinfo {author} {\bibfnamefont {I.}~\bibnamefont {Giardina}},\
  and\ \bibinfo {author} {\bibfnamefont {G.}~\bibnamefont {Parisi}},\ }\href
  {https://doi.org/10.1103/PhysRevB.57.11251} {\bibfield  {journal} {\bibinfo
  {journal} {Phys. Rev. B}\ }\textbf {\bibinfo {volume} {57}},\ \bibinfo
  {pages} {11251} (\bibinfo {year} {1998})}\BibitemShut {NoStop}%
\bibitem [{\citenamefont {Baxter}(1971)}]{Baxter1971}%
  \BibitemOpen
  \bibfield  {author} {\bibinfo {author} {\bibfnamefont {R.~J.}\ \bibnamefont
  {Baxter}},\ }\href {https://doi.org/10.1103/PhysRevLett.26.832} {\bibfield
  {journal} {\bibinfo  {journal} {Phys. Rev. Lett.}\ }\textbf {\bibinfo
  {volume} {26}},\ \bibinfo {pages} {832} (\bibinfo {year} {1971})}\BibitemShut
  {NoStop}%
\bibitem [{\citenamefont {Kadanoff}\ and\ \citenamefont
  {Wegner}(1971)}]{Kadanoff1971}%
  \BibitemOpen
  \bibfield  {author} {\bibinfo {author} {\bibfnamefont {L.~P.}\ \bibnamefont
  {Kadanoff}}\ and\ \bibinfo {author} {\bibfnamefont {F.~J.}\ \bibnamefont
  {Wegner}},\ }\href {https://doi.org/10.1103/PhysRevB.4.3989} {\bibfield
  {journal} {\bibinfo  {journal} {Phys. Rev. B}\ }\textbf {\bibinfo {volume}
  {4}},\ \bibinfo {pages} {3989} (\bibinfo {year} {1971})}\BibitemShut
  {NoStop}%
\bibitem [{\citenamefont {Bernardi}\ and\ \citenamefont
  {Campbell}(1995)}]{Bernardi1995}%
  \BibitemOpen
  \bibfield  {author} {\bibinfo {author} {\bibfnamefont {L.}~\bibnamefont
  {Bernardi}}\ and\ \bibinfo {author} {\bibfnamefont {I.~A.}\ \bibnamefont
  {Campbell}},\ }\href {https://doi.org/10.1103/PhysRevB.52.12501} {\bibfield
  {journal} {\bibinfo  {journal} {Phys. Rev. B}\ }\textbf {\bibinfo {volume}
  {52}},\ \bibinfo {pages} {12501} (\bibinfo {year} {1995})}\BibitemShut
  {NoStop}%
\bibitem [{\citenamefont {Jin}\ \emph {et~al.}(2012)\citenamefont {Jin},
  \citenamefont {Sen},\ and\ \citenamefont {Sandvik}}]{Jin2012}%
  \BibitemOpen
  \bibfield  {author} {\bibinfo {author} {\bibfnamefont {S.}~\bibnamefont
  {Jin}}, \bibinfo {author} {\bibfnamefont {A.}~\bibnamefont {Sen}},\ and\
  \bibinfo {author} {\bibfnamefont {A.~W.}\ \bibnamefont {Sandvik}},\ }\href
  {https://doi.org/10.1103/PhysRevLett.108.045702} {\bibfield  {journal}
  {\bibinfo  {journal} {Phys. Rev. Lett.}\ }\textbf {\bibinfo {volume} {108}},\
  \bibinfo {pages} {045702} (\bibinfo {year} {2012})}\BibitemShut {NoStop}%
\bibitem [{\citenamefont {Lang}\ \emph {et~al.}(2024)\citenamefont {Lang},
  \citenamefont {Sachdev},\ and\ \citenamefont {Diehl}}]{Lang2024b}%
  \BibitemOpen
  \bibfield  {author} {\bibinfo {author} {\bibfnamefont {J.}~\bibnamefont
  {Lang}}, \bibinfo {author} {\bibfnamefont {S.}~\bibnamefont {Sachdev}},\ and\
  \bibinfo {author} {\bibfnamefont {S.}~\bibnamefont {Diehl}},\ }\href
  {https://doi.org/10.21468/SciPostPhys.17.6.160} {\bibfield  {journal}
  {\bibinfo  {journal} {SciPost Phys.}\ }\textbf {\bibinfo {volume} {17}},\
  \bibinfo {pages} {160} (\bibinfo {year} {2024})}\BibitemShut {NoStop}%
\bibitem [{\citenamefont {Ketcheson}\ \emph {et~al.}(2009)\citenamefont
  {Ketcheson}, \citenamefont {Macdonald},\ and\ \citenamefont
  {Gottlieb}}]{Ketcheson2009}%
  \BibitemOpen
  \bibfield  {author} {\bibinfo {author} {\bibfnamefont {D.~I.}\ \bibnamefont
  {Ketcheson}}, \bibinfo {author} {\bibfnamefont {C.~B.}\ \bibnamefont
  {Macdonald}},\ and\ \bibinfo {author} {\bibfnamefont {S.}~\bibnamefont
  {Gottlieb}},\ }\href
  {https://doi.org/https://doi.org/10.1016/j.apnum.2008.03.034} {\bibfield
  {journal} {\bibinfo  {journal} {Applied Numerical Mathematics}\ }\textbf
  {\bibinfo {volume} {59}},\ \bibinfo {pages} {373} (\bibinfo {year}
  {2009})}\BibitemShut {NoStop}%
\bibitem [{\citenamefont {Izzo}\ and\ \citenamefont
  {Jackiewicz}(2022)}]{Izzo2022}%
  \BibitemOpen
  \bibfield  {author} {\bibinfo {author} {\bibfnamefont {G.}~\bibnamefont
  {Izzo}}\ and\ \bibinfo {author} {\bibfnamefont {Z.}~\bibnamefont
  {Jackiewicz}},\ }\href {https://doi.org/10.1007/s41980-022-00731-x}
  {\bibfield  {journal} {\bibinfo  {journal} {Bulletin of the Iranian
  Mathematical Society}\ }\textbf {\bibinfo {volume} {48}},\ \bibinfo {pages}
  {4029} (\bibinfo {year} {2022})}\BibitemShut {NoStop}%
\bibitem [{\citenamefont {Bernaschi}\ \emph {et~al.}(2020)\citenamefont
  {Bernaschi}, \citenamefont {Billoire}, \citenamefont {Maiorano},
  \citenamefont {Parisi},\ and\ \citenamefont
  {Ricci-Tersenghi}}]{Bernaschi2020}%
  \BibitemOpen
  \bibfield  {author} {\bibinfo {author} {\bibfnamefont {M.}~\bibnamefont
  {Bernaschi}}, \bibinfo {author} {\bibfnamefont {A.}~\bibnamefont {Billoire}},
  \bibinfo {author} {\bibfnamefont {A.}~\bibnamefont {Maiorano}}, \bibinfo
  {author} {\bibfnamefont {G.}~\bibnamefont {Parisi}},\ and\ \bibinfo {author}
  {\bibfnamefont {F.}~\bibnamefont {Ricci-Tersenghi}},\ }\href
  {https://doi.org/10.1073/pnas.1910936117} {\bibfield  {journal} {\bibinfo
  {journal} {Proceedings of the National Academy of Sciences}\ }\textbf
  {\bibinfo {volume} {117}},\ \bibinfo {pages} {17522} (\bibinfo {year}
  {2020})},\ \eprint
  {https://www.pnas.org/doi/pdf/10.1073/pnas.1910936117} \BibitemShut {NoStop}%
\bibitem [{\citenamefont {Kennett}\ \emph {et~al.}(2001)\citenamefont
  {Kennett}, \citenamefont {Chamon},\ and\ \citenamefont {Ye}}]{Kennett2001b}%
  \BibitemOpen
  \bibfield  {author} {\bibinfo {author} {\bibfnamefont {M.~P.}\ \bibnamefont
  {Kennett}}, \bibinfo {author} {\bibfnamefont {C.}~\bibnamefont {Chamon}},\
  and\ \bibinfo {author} {\bibfnamefont {J.}~\bibnamefont {Ye}},\ }\href
  {https://doi.org/10.1103/PhysRevB.64.224408} {\bibfield  {journal} {\bibinfo
  {journal} {Phys. Rev. B}\ }\textbf {\bibinfo {volume} {64}},\ \bibinfo
  {pages} {224408} (\bibinfo {year} {2001})}\BibitemShut {NoStop}%
\bibitem [{\citenamefont {{Gardner, E.}}\ \emph {et~al.}(1987)\citenamefont
  {{Gardner, E.}}, \citenamefont {{Derrida, B.}},\ and\ \citenamefont
  {{Mottishaw, P.}}}]{Gardner1987}%
  \BibitemOpen
  \bibfield  {author} {\bibinfo {author} {\bibnamefont {{Gardner, E.}}},
  \bibinfo {author} {\bibnamefont {{Derrida, B.}}},\ and\ \bibinfo {author}
  {\bibnamefont {{Mottishaw, P.}}},\ }\href
  {https://doi.org/10.1051/jphys:01987004805074100} {\bibfield  {journal}
  {\bibinfo  {journal} {J. Phys. France}\ }\textbf {\bibinfo {volume} {48}},\
  \bibinfo {pages} {741} (\bibinfo {year} {1987})}\BibitemShut {NoStop}%
\bibitem [{\citenamefont {Bray}\ and\ \citenamefont {Dean}(2007)}]{Dean2007}%
  \BibitemOpen
  \bibfield  {author} {\bibinfo {author} {\bibfnamefont {A.~J.}\ \bibnamefont
  {Bray}}\ and\ \bibinfo {author} {\bibfnamefont {D.~S.}\ \bibnamefont
  {Dean}},\ }\href {https://doi.org/10.1103/PhysRevLett.98.150201} {\bibfield
  {journal} {\bibinfo  {journal} {Phys. Rev. Lett.}\ }\textbf {\bibinfo
  {volume} {98}},\ \bibinfo {pages} {150201} (\bibinfo {year}
  {2007})}\BibitemShut {NoStop}%
\bibitem [{\citenamefont {Baity-Jesi}\ \emph {et~al.}(2019)\citenamefont
  {Baity-Jesi}, \citenamefont {Sagun}, \citenamefont {Geiger}, \citenamefont
  {Spigler}, \citenamefont {Ben~Arous}, \citenamefont {Cammarota},
  \citenamefont {LeCun}, \citenamefont {Wyart},\ and\ \citenamefont
  {Biroli}}]{Baity-Jesi2019}%
  \BibitemOpen
  \bibfield  {author} {\bibinfo {author} {\bibfnamefont {M.}~\bibnamefont
  {Baity-Jesi}}, \bibinfo {author} {\bibfnamefont {L.}~\bibnamefont {Sagun}},
  \bibinfo {author} {\bibfnamefont {M.}~\bibnamefont {Geiger}}, \bibinfo
  {author} {\bibfnamefont {S.}~\bibnamefont {Spigler}}, \bibinfo {author}
  {\bibfnamefont {G.}~\bibnamefont {Ben~Arous}}, \bibinfo {author}
  {\bibfnamefont {C.}~\bibnamefont {Cammarota}}, \bibinfo {author}
  {\bibfnamefont {Y.}~\bibnamefont {LeCun}}, \bibinfo {author} {\bibfnamefont
  {M.}~\bibnamefont {Wyart}},\ and\ \bibinfo {author} {\bibfnamefont
  {G.}~\bibnamefont {Biroli}},\ }\href
  {https://doi.org/10.1088/1742-5468/ab3281} {\bibfield  {journal} {\bibinfo
  {journal} {Journal of Statistical Mechanics: Theory and Experiment}\ }\textbf
  {\bibinfo {volume} {2019}},\ \bibinfo {pages} {124013} (\bibinfo {year}
  {2019})}\BibitemShut {NoStop}%
\bibitem [{\citenamefont {Citro}\ and\ \citenamefont
  {Ricci-Tersenghi}(2025)}]{Citro2025}%
  \BibitemOpen
  \bibfield  {author} {\bibinfo {author} {\bibfnamefont {V.}~\bibnamefont
  {Citro}}\ and\ \bibinfo {author} {\bibfnamefont {F.}~\bibnamefont
  {Ricci-Tersenghi}},\ }\href {https://link.aps.org/doi/10.1103/lvmp-pydk}
  {\bibfield  {journal} {\bibinfo  {journal} {Phys. Rev. Lett.}\ }\textbf
  {\bibinfo {volume} {135}},\ \bibinfo {pages} {247102} (\bibinfo {year}
  {2025})}\BibitemShut {NoStop}%
\end{thebibliography}

\begin{thebibliography}{12}
\expandafter\ifx\csname natexlab\endcsname\relax\def\natexlab#1{#1}\fi
\expandafter\ifx\csname bibnamefont\endcsname\relax
  \def\bibnamefont#1{#1}\fi
\expandafter\ifx\csname bibfnamefont\endcsname\relax
  \def\bibfnamefont#1{#1}\fi
\expandafter\ifx\csname citenamefont\endcsname\relax
  \def\citenamefont#1{#1}\fi
\expandafter\ifx\csname url\endcsname\relax
  \def\url#1{\texttt{#1}}\fi
\expandafter\ifx\csname urlprefix\endcsname\relax\def\urlprefix{URL }\fi
\providecommand{\bibinfo}[2]{#2}
\providecommand{\eprint}[2][]{\url{#2}}

\bibitem [{\citenamefont {Gustafsson}(1994)}]{Gustafsson1994S}%
  \BibitemOpen
  \bibfield  {author} {\bibinfo {author} {\bibfnamefont {K.}~\bibnamefont
  {Gustafsson}},\ }\href {https://doi.org/10.1145/198429.198437} {\bibfield
  {journal} {\bibinfo  {journal} {ACM Trans. Math. Softw.}\ }\textbf {\bibinfo
  {volume} {20}},\ \bibinfo {pages} {496–517} (\bibinfo {year}
  {1994})}\BibitemShut {NoStop}%
\bibitem [{\citenamefont {Gustafsson}\ and\ \citenamefont
  {S\"{o}derlind}(1997)}]{Gustafsson1997S}%
  \BibitemOpen
  \bibfield  {author} {\bibinfo {author} {\bibfnamefont {K.}~\bibnamefont
  {Gustafsson}}\ and\ \bibinfo {author} {\bibfnamefont {G.}~\bibnamefont
  {S\"{o}derlind}},\ }\href {https://doi.org/10.1137/S1064827595287109}
  {\bibfield  {journal} {\bibinfo  {journal} {SIAM Journal on Scientific
  Computing}\ }\textbf {\bibinfo {volume} {18}},\ \bibinfo {pages} {23}
  (\bibinfo {year} {1997})},\ \eprint
  {https://doi.org/10.1137/S1064827595287109} \BibitemShut {NoStop}%
\bibitem [{\citenamefont {Burrage}\ \emph {et~al.}(2004)\citenamefont
  {Burrage}, \citenamefont {Herdiana},\ and\ \citenamefont
  {Burrage}}]{Burrage2004S}%
  \BibitemOpen
  \bibfield  {author} {\bibinfo {author} {\bibfnamefont {P.}~\bibnamefont
  {Burrage}}, \bibinfo {author} {\bibfnamefont {R.}~\bibnamefont {Herdiana}},\
  and\ \bibinfo {author} {\bibfnamefont {K.}~\bibnamefont {Burrage}},\ }\href
  {https://doi.org/https://doi.org/10.1016/j.cam.2004.01.027} {\bibfield
  {journal} {\bibinfo  {journal} {Journal of Computational and Applied
  Mathematics}\ }\textbf {\bibinfo {volume} {170}},\ \bibinfo {pages} {317}
  (\bibinfo {year} {2004})}\BibitemShut {NoStop}%
\bibitem [{\citenamefont {Dormand}\ and\ \citenamefont
  {Prince}(1980)}]{Dormand1980S}%
  \BibitemOpen
  \bibfield  {author} {\bibinfo {author} {\bibfnamefont {J.}~\bibnamefont
  {Dormand}}\ and\ \bibinfo {author} {\bibfnamefont {P.}~\bibnamefont
  {Prince}},\ }\href
  {https://doi.org/https://doi.org/10.1016/0771-050X(80)90013-3} {\bibfield
  {journal} {\bibinfo  {journal} {Journal of Computational and Applied
  Mathematics}\ }\textbf {\bibinfo {volume} {6}},\ \bibinfo {pages} {19}
  (\bibinfo {year} {1980})}\BibitemShut {NoStop}%
\bibitem [{\citenamefont {Martín-Vaquero}\ and\ \citenamefont
  {Janssen}(2009)}]{Vaquero2009S}%
  \BibitemOpen
  \bibfield  {author} {\bibinfo {author} {\bibfnamefont {J.}~\bibnamefont
  {Martín-Vaquero}}\ and\ \bibinfo {author} {\bibfnamefont {B.}~\bibnamefont
  {Janssen}},\ }\href
  {https://doi.org/https://doi.org/10.1016/j.cpc.2009.05.006} {\bibfield
  {journal} {\bibinfo  {journal} {Computer Physics Communications}\ }\textbf
  {\bibinfo {volume} {180}},\ \bibinfo {pages} {1802} (\bibinfo {year}
  {2009})}\BibitemShut {NoStop}%
\bibitem [{\citenamefont {Martín-Vaquero}\ and\ \citenamefont
  {Kleefeld}(2016)}]{Vaquero2016S}%
  \BibitemOpen
  \bibfield  {author} {\bibinfo {author} {\bibfnamefont {J.}~\bibnamefont
  {Martín-Vaquero}}\ and\ \bibinfo {author} {\bibfnamefont {B.}~\bibnamefont
  {Kleefeld}},\ }\href
  {https://doi.org/https://doi.org/10.1016/j.jcp.2016.08.042} {\bibfield
  {journal} {\bibinfo  {journal} {Journal of Computational Physics}\ }\textbf
  {\bibinfo {volume} {326}},\ \bibinfo {pages} {141} (\bibinfo {year}
  {2016})}\BibitemShut {NoStop}%
\bibitem [{\citenamefont {Abdulle}(2002)}]{Abdulle2002S}%
  \BibitemOpen
  \bibfield  {author} {\bibinfo {author} {\bibfnamefont {A.}~\bibnamefont
  {Abdulle}},\ }\href {https://doi.org/10.1137/S1064827500379549} {\bibfield
  {journal} {\bibinfo  {journal} {SIAM Journal on Scientific Computing}\
  }\textbf {\bibinfo {volume} {23}},\ \bibinfo {pages} {2041} (\bibinfo {year}
  {2002})},\ \eprint
  {https://doi.org/10.1137/S1064827500379549} \BibitemShut {NoStop}%
\bibitem [{\citenamefont {Cugliandolo}\ and\ \citenamefont
  {Kurchan}(1994)}]{Cugliandolo1994S}%
  \BibitemOpen
  \bibfield  {author} {\bibinfo {author} {\bibfnamefont {L.~F.}\ \bibnamefont
  {Cugliandolo}}\ and\ \bibinfo {author} {\bibfnamefont {J.}~\bibnamefont
  {Kurchan}},\ }\href {https://doi.org/10.1088/0305-4470/27/17/011} {\bibfield
  {journal} {\bibinfo  {journal} {Journal of Physics A: Mathematical and
  General}\ }\textbf {\bibinfo {volume} {27}},\ \bibinfo {pages} {5749}
  (\bibinfo {year} {1994})}\BibitemShut {NoStop}%
\bibitem [{\citenamefont {Cugliandolo}\ and\ \citenamefont
  {Kurchan}(1995)}]{Cugliandolo1995S}%
  \BibitemOpen
  \bibfield  {author} {\bibinfo {author} {\bibfnamefont {L.~F.}\ \bibnamefont
  {Cugliandolo}}\ and\ \bibinfo {author} {\bibfnamefont {J.}~\bibnamefont
  {Kurchan}},\ }\href {https://doi.org/10.1080/01418639508238541} {\bibfield
  {journal} {\bibinfo  {journal} {Philosophical Magazine B}\ }\textbf {\bibinfo
  {volume} {71}},\ \bibinfo {pages} {501} (\bibinfo {year} {1995})}\BibitemShut
  {NoStop}%
\bibitem [{\citenamefont {Cugliandolo}\ and\ \citenamefont
  {Kurchan}(1993)}]{Cugliandolo1993S}%
  \BibitemOpen
  \bibfield  {author} {\bibinfo {author} {\bibfnamefont {L.~F.}\ \bibnamefont
  {Cugliandolo}}\ and\ \bibinfo {author} {\bibfnamefont {J.}~\bibnamefont
  {Kurchan}},\ }\href {https://doi.org/10.1103/PhysRevLett.71.173} {\bibfield
  {journal} {\bibinfo  {journal} {Phys. Rev. Lett.}\ }\textbf {\bibinfo
  {volume} {71}},\ \bibinfo {pages} {173} (\bibinfo {year} {1993})}\BibitemShut
  {NoStop}%
\bibitem [{\citenamefont {Folena}\ \emph {et~al.}(2020)\citenamefont {Folena},
  \citenamefont {Franz},\ and\ \citenamefont {Ricci-Tersenghi}}]{Folena2020S}%
  \BibitemOpen
  \bibfield  {author} {\bibinfo {author} {\bibfnamefont {G.}~\bibnamefont
  {Folena}}, \bibinfo {author} {\bibfnamefont {S.}~\bibnamefont {Franz}},\ and\
  \bibinfo {author} {\bibfnamefont {F.}~\bibnamefont {Ricci-Tersenghi}},\
  }\href {https://doi.org/10.1103/PhysRevX.10.031045} {\bibfield  {journal}
  {\bibinfo  {journal} {Phys. Rev. X}\ }\textbf {\bibinfo {volume} {10}},\
  \bibinfo {pages} {031045} (\bibinfo {year} {2020})}\BibitemShut {NoStop}%
\bibitem [{\citenamefont {Folena}\ and\ \citenamefont
  {Zamponi}(2023)}]{Folena2023S}%
  \BibitemOpen
  \bibfield  {author} {\bibinfo {author} {\bibfnamefont {G.}~\bibnamefont
  {Folena}}\ and\ \bibinfo {author} {\bibfnamefont {F.}~\bibnamefont
  {Zamponi}},\ }\href {https://doi.org/10.21468/SciPostPhys.15.3.109}
  {\bibfield  {journal} {\bibinfo  {journal} {SciPost Phys.}\ }\textbf
  {\bibinfo {volume} {15}},\ \bibinfo {pages} {109} (\bibinfo {year}
  {2023})}
\end{thebibliography}
\end{document}